\documentclass[11pt]{article}      
\pdfoutput=1
\usepackage{amsthm,amsmath,amssymb,xcolor,sectsty,latexsym,mathtools,hyperref,mathrsfs}
\usepackage[shortlabels]{enumitem}
\usepackage{slashed}  
\usepackage{bm}
\usepackage{epsfig}
\usepackage{dcolumn}
\definecolor{vermelho}{cmyk}{0,.88,.77,.40}
\usepackage{cite}
\usepackage{feynmp}
\numberwithin{equation}{section}
\usepackage[textwidth=6.5in,top=1.2in,bottom=1.2in]{geometry}
\usepackage{setspace}
\chapterfont{\color{vermelho}}
\usepackage{simplewick}
\usepackage{hyperref}
\usepackage{float}
\usepackage{mathtools}
\usepackage{blkarray, bigstrut} %
\usepackage{graphicx}

\newcommand{\be}{\begin{equation}}
\newcommand{\ee}{\end{equation}}
\newcommand{\beq}{\begin{equation}}
\newcommand{\eeq}{\end{equation}}
\newcommand{\ba}{\begin{eqnarray}}
\newcommand{\ea}{\end{eqnarray}}
\newcommand{\bea}{\begin{eqnarray}}
\newcommand{\eea}{\end{eqnarray}}
\newcommand{\bef}{\begin{figure}}
\newcommand{\eef}{\end{figure}}

\newcommand{\mpl}{M_{\mbox{\tiny{Pl}}}}
\newcommand{\T}{\mu}
\newcommand{\rhoeff}{\rho_{\mbox{\tiny{eff}}}}
\newcommand{\peff}{\mathcal{P}_{\mbox{\tiny{eff}}}}
\newcommand{\pert}{\varepsilon}
\newcommand{\norm}{\mathcal{N}}
\newcommand{\Bs}{B_\phi}
\newcommand{\redu}{m}
\newcommand{\Per}{\mathcal{T}}

\begin{document}

\thispagestyle{empty}
\begin{titlepage}
\nopagebreak

\title{  \begin{center}\bf Causal Modifications of Gravity \\ and Their Observational Bounds \\\end{center} }

\vfill
\author{Mark P.~Hertzberg$^1$\footnote{mark.hertzberg@tufts.edu}, Jacob A.~Litterer$^{1,2}$\footnote{jacob.litterer@tufts.edu}, Neil Shah$^{1}$\footnote{neil.shah@tufts.edu}}
\date{ }

\maketitle

\begin{center}
	\vspace{-0.7cm}
	{\it  $^1$Institute of Cosmology, Department of Physics and Astronomy}\\
	{\it  Tufts University, Medford, MA 02155, USA}\\
	{\it $^2$Univ Coimbra, Faculdade de Ci\^encias e Tecnologia
	da Universidade \\ de Coimbra and CFisUC, Rua Larga, 3004-516 Coimbra, Portugal}
	\end{center}
\bigskip

\begin{abstract}
Since general relativity is the unique theory of massless spin 2 particles at large distances, the most reasonable way to have significant modifications is to introduce one or more light scalars that mediate a new long-range force. Most existing studies of such scalars invoke models that exhibit some kind of ``screening" at short distances to hide the force from solar system tests. However, as is well known, such modifications also exhibit superluminality, which can be interpreted as a form of acausality. In this work we explore explicitly subluminal and causal scalar field models. In particular, we study a conformally coupled scalar $\phi$, with a small coupling to matter to obey solar system bounds, and a non-canonical kinetic term $K(X)$ ($X=(\partial\phi)^2/2$) that obeys all subluminality constraints and is hyperbolic. We consider $K(X)$ that is canonical for small $X$, but beyond some nonlinear scale enters a new scaling regime of power $p$, with $1/2<p<1$ (the DBI kinetic term is the limit $p=1/2$ and a canonical scalar is $p=1$). As opposed to screening (and superluminality), this new force becomes more and more important in the regime of high densities (and subluminality). We then turn to the densest environments to put bounds on this new interaction. We compute constraints from precession in binary systems such as Hulse-Taylor, we compute corrections to neutron star hydrostatic equilibrium, and we compute power in radiation, both tensor mode corrections and the new scalar mode, which can be important during mergers. 
\end{abstract}

\end{titlepage}

\setcounter{page}{2}

\tableofcontents

\newpage

\section{Introduction}
It is interesting to explore potential deviations in gravity from general relativity. However, building theoretically consistent alternatives is very difficult. This is because general relativity is the unique Lorentz invariant, causal, theory of massless spin 2 particles at large distances \cite{Weinberg:1964ew,Weinberg:1965rz,Deser:1969wk,Feynman,Hertzberg:2016djj,Hertzberg:2017abn} (and even the assumption of Lorentz invariance may be partially relaxed in favor of only assuming rotations and locality \cite{Khoury:2013oqa,Hertzberg:2017nzl,Hertzberg:2020yzl,Hertzberg:2020gxu}.) The only option then to obtain anything new within the framework of special relativity is to introduce some other representations of the Lorentz group. Generically, these cannot be fermions, as they cannot mediate long-ranged forces. It is also quite difficult to make use of a new spin 1 particle because it necessarily makes opposites attract and likes repels, so systems tend to find ways to neutralize this force, as they do for standard electromagnetism. Therefore, the only reasonable possibility is to introduce one or more spinless particles. If these particles are sufficiently light, they may give rise to a new long-range attractive force. This acts as a plausible form of modified gravity. The lightness of such particles is not, however, protected by any obvious consideration. We cannot appeal to degree of freedom counting, as we can for spin $s\geq1$, nor can we appeal to chiral charge assignments, as we can for spin $s=1/2$, to explain its lightness. Nevertheless, scalars that are assumed to couple universally are somewhat robust against quantum corrections, as we will recap in the next section.

However, standard scalar couplings, at first sight seem to be immediately ruled out by solar systems tests, which require any new force to be several orders of magnitude weaker than gravity. To avoid this, many authors have appealed to screening mechanisms that hide the new force in high density environments, by transitioning from a $\propto 1/r^2$ force law to a $\propto 1/r^{2+q}$ force law with $q<0$. Such models include massive gravity \cite{deRham:2010kj}, Galileon models \cite{deRham:2012az}, etc, that may be relevant to dark energy, or MOND \cite{Milgrom:1983ca,Milgrom:1983pn,Milgrom:1983zz,Bekenstein:1984tv,Bekenstein:2004ne,Bruneton:2008fk,Blanchet:2006yt,Bernard:2014psa,Skordis:2020eui}, superfluid \cite{Berezhiani:2015pia,Berezhiani:2015bqa,Khoury:2016ehj,Ferreira:2018wup} models, etc, that may be relevant to dark matter. However, as we will recap in the next section, in a Lorentz invariant theory, there is a close relationship between the scaling of the force and the speed of high energy perturbations. In particular, if the force is {\em slower} that standard $1/r^2$, there is superluminality. So if we invoke the absence of superluminality as a principle, then we must exclude these models. In fact it is believed that such low energy effective field theories cannot possess any Lorentz invariant Wilsonian UV completion. Superluminality, or another issue of breaking of hyperbolicity, we shall refer to as forms of acausality (see ahead to Section \ref{HypSuperConstraints}).

On the other hand, this same argument lends itself to a new interesting possibility: what if $q>0$ and the force rises {\em faster} than standard $1/r^2$ in high density environments? Instead these give rise to healthy subluminal perturbations, and hence, may conceivably possess a sensible UV completion. By invoking a small coupling to matter, we can trivially evade the solar system bounds. This may not be so relevant to dark energy or dark matter. But what is exciting instead is that the new force can transition to a new regime in which it starts to compete with general relativity in very high density environments. This is our new idea that we will explore in this work. In particular, we will be interested in neutron star systems, which are some of the highest density environments in the universe. (We will also comment on black holes). We will begin by studying leading order corrections to orbital motion, finding an important correction to the precession of systems such as the Hulse-Taylor binary. We then move to consider the modified structure of neutron stars. And we study scalar radiation from mergers. Our primary constraints are summarized ahead in Fig.~\ref{fig:ConstraintPlot}.

Our paper is organized as follows: 
in Section~\ref{sec:Models}, we introduce this new class of models, standard solar system bounds, and develop basic theoretical constraints.
In Section~\ref{sec:Orbital}, we compute corrections to orbital motion from the new scalar force, applying this to the Hulse-Taylor system.
In Section~\ref{sec:Equilibrium}, we establish the modified Tolman-Oppenheimer-Volkoff  equations for neutron star equilibrium and solve them numerically.
In Section~\ref{sec:Emission}, we analytically compute the scalar wave emission in both the linear and nonlinear regimes and corrections to tensor mode power, and comment on mergers.
Finally, in Section~\ref{sec:Discussion}, we discuss our findings and mention possible future directions.

\section{A Class of Causal Scalar Field Models}\label{sec:Models}

In this work we introduce a scalar $\phi$ to mediate a new long-range force. We know that a generic scalar with non-universal  couplings to matter will lead to a breakdown of universal free-fall (weak equivalence principle) and is ruled out observationally.  However, generic non-universally coupled scalars are not associated with any symmetry and so are expected to be quite massive anyhow, and therefore irrelevant at large distances. 

Instead let us suppose there exists a scalar that happens to be universally coupled like the graviton. While the graviton's universal coupling is guaranteed by Lorentz symmetry, a scalar's universal coupling is evidently not. Nevertheless an assumed universal coupling is surprisingly stable against renormalization. In particular any matter loops do not alter this coupling \cite{Hui:2010dn}. Corrections from loops can generate a mass for $\phi$, though that will be at least Planck suppressed. 
In particular, a matter particle running in a loop will tend to induce a mass $m_\phi\sim c\, \Lambda^2/(4\pi\mpl)$. Moreover, for a conformally coupled scalar, the mass is in some sense tied to the generation of the cosmological constant. So by tuning that to be small, we should work to next order, giving the scalar-matter loop contribution of  \cite{Hertzberg:2018suv} 
$m_\phi \sim c^2 \Lambda^3/((4\pi)^2\mpl^2)$ 
where $\Lambda$ is the cutoff on the effective theory, 
$\mpl=1/\sqrt{8\pi G}$ is the (reduced) Planck mass, and $c$ is a dimensionless coupling (shortly we replace notation $c\to\beta\,\mpl$). 
Of course if the cutoff is pushed towards the Planck scale, then one expects a heavy scalar. However, for a low cutoff, the Planck suppression can lead to a rather light scalar. Such a scalar could therefore potentially give rise to a new long-range force. So a universally coupled scalar (or ``conformally coupled", as we will discuss) is potentially interesting.

Let us then build an action for a very light scalar $\phi$ coupled to (standard model) matter $\psi_i$ and gravity $g_{\mu\nu}$. The graviton must be minimally coupled (at leading order) of course, and the universal coupling of the scalar is similarly implemented by a rescaling of the effective metric. So let us define a kind of effective metric
\beq
\tilde{g}_{\mu\nu}\equiv f(\phi)\,g_{\mu\nu}
\eeq
where $f(\phi)$ is a function that we can choose. By coupling to matter this way, we see that it is the same as saying matter is moving in a new metric $\tilde{g}_{\mu\nu}$ that is conformally related to the original metric $g_{\mu\nu}$, and hence this is a ``conformally coupled" scalar.

We shall assume without loss of generality that the vacuum of the theory is at $\phi=0$, and here we can take $f(0)=1$, recovering a standard metric. We shall also assume that for small $\phi$, $f(\phi)$ has a linear correction
\beq
f(\phi)=1-2\,\beta\,\phi+\ldots
\eeq
Such a linear term will lead to $\phi$ mediating a long-ranged force. The parameter $\beta$, which has units of inverse mass, sets the strength of the coupling. As we shall see, solar system constraints will demand that $\beta$ is smaller than $1/\mpl$. Since $f(\phi)$ represents a breaking of a shift symmetry for $\phi$, it is technically natural for $\beta$ to be arbitrarily small (later we shall discuss this further).

The full action in this work is taken to be (signature + - - -, with $c=\hbar=1$)
\beq
S=\int d^4x\sqrt{-g}\left[{\mathcal{R}\over 16\pi G}+f(\phi)^2\mathcal{L}_M(\psi_i,\tilde{g}_{\mu\nu})+K(X)\right]
\eeq
The first term is the standard Einstein Hilbert term, the second term is the matter Lagrangian with its conformal coupling to $\phi$, via $\tilde{g}_{\mu\nu}=f(\phi)g_{\mu\nu}$. We shall assume that $\mathcal{L}_M$ includes the Standard Model Lagrangian, though it may include dark matter as well (dark matter will not play a central role in the present work). 

Finally, and very importantly, we have introduced a kinetic term for the scalar $K(X)$, where
\beq
X\equiv {1\over 2}g^{\mu\nu}\partial_\mu\phi\partial_\nu\phi
\eeq
is the Lorentz invariant kinetic term. In order to obtain some novel dynamics, we need to consider non-canonical choices for $K(X)$, which we will discuss shortly. Applications of non-standard kinetic terms to dark energy, inflation, and fifth forces include Refs.~\cite{Garriga:1999vw,ArmendarizPicon:1999rj,Rendall:2005fv,Brax:2012jr,Brax:2014wla,Barreira:2015aea,Bains:2015gpv,Bezares:2020wkn}.

\subsection{Standard Solar System Bound}\label{SSB}

To set the stage, let us note that if we pick the canonical value 
\beq
K(X)=X
\eeq
then we know that $\phi$ will simply lead to a standard $\propto 1/r^2$ force law (between non-relativistic matter). Such a new force is constrained by solar system tests of gravity, requiring the coupling $\beta$ to be much smaller than $1/\mpl$, the spin 2 coupling. More precisely, the equation of motion for a canonical $\phi$ is
\beq
\Box\phi=-T_M{f'(\phi)\over 2\,f(\phi)}
\eeq
where $f'\equiv df/d\phi$ and the box operator here is standard for a scalar in curved spacetime, i,e.,
\beq
\Box={1\over\sqrt{-g}}\,\partial_\mu\left(\sqrt{-g}\,g^{\mu\alpha}\partial_\alpha\phi\right)
\eeq
and $T_M$ is the trace of the energy momentum tensor of the matter ``M" sector
\beq
T_{M}=g^{\mu\nu}T^{(M)}_{\mu\nu}
\eeq
with $T^{(M)}_{\mu\nu}$ computed from the {\em full} matter Lagrangian including the conformal factor $f(\phi)$ (i.e., $T^{(M)}_{\mu\nu}$ can depend on $\phi$ also). 

In the weak field regime relevant to the solar system, such differences are unimportant. We have to leading order $f(\phi)\approx 1$, $f'(\phi)\approx-2\beta$ and $T_{M}\approx\rho-3\mathcal{P}$, where $\rho$ and $\mathcal{P}$ are the energy density and pressure of matter computed in the absence of $\phi$, respectively. This gives
\beq
\Box\phi\approx\beta\,(\rho-3\mathcal{P})
\eeq
Since the right hand side vanishes for light, we know that $\phi$ does not lead to any light bending, and this puts a serious observational constraint on these models. To leading approximation, the value of $\phi$ from the sun and planets in the solar system is then obtained by solving the static limit of the above equation (and ignoring $\mathcal{P}$), which is just the Poisson equation $-\nabla^2\phi=\beta\rho$. We can compare this to the Newton potential $\phi_N$ in the Newtonian gauge, which obeys $\nabla^2\phi_N=4\pi G\rho=\rho/(2\mpl^2)$, yielding
\beq
\phi\approx -2\beta\mpl^2\,\phi_N
\eeq 
So in the solar system, $\phi$ and the Newton potential $\phi_N$ are proportional to each other, with a constant of proportionality $-2\beta\,\mpl^2$.
Then a convenient way to calculate the consequence for light is to simply return to the fact that all matter is now coupled to the effective metric $\tilde{g}_{\mu\nu}\approx g_{\mu\nu}(1-2\beta\phi)$, which can then be expanded as \cite{SaksteinNotes}
\beq
\tilde{g}_{\mu\nu}\approx(1+2\phi_N(1+2\beta^2\mpl^2))dt^2-(1-2\phi_N(1-2\beta^2\mpl^2))|d{\bf x}|^2
\eeq
Since we fix the Newton's constant $G$ to match non-relativistic behavior, then the ratio of these corrections represent an effective alteration in the bending of light $\delta\theta$ compared to general relativity $\theta_{GR}=4GM/b$. To leading order the relative correction is therefore 
\beq
{\delta\theta\over \theta_{GR}}=4\beta^2\mpl^2
\eeq
The tightest bound on this comes from the Cassini probe of
\beq
{\delta\theta\over \theta_{GR}}\leq 2.5\times 10^{-5}
\eeq
Hence we obtain a (well known) bound on the scalar coupling of
\beq
\beta\leq{2.5\times 10^{-3}\over\mpl}
\label{betaconstraint}\eeq
Such a small coupling may at first sight seem ``un-natural" or ``fine-tuned". But from the effective field theory point of view it is not. Since $\beta$ introduces a soft breaking of a shift symmetry in $\phi$, it is technically natural for it to be arbitrarily small, i.e., in the limit as $\beta\to 0$, the shift symmetry is recovered and it will not be generated. So any small, but non-zero, $\beta$ is very stable against renormalization. On the other hand, whether such a small $\beta$ is reasonable from the point of view of quantum gravity is less clear; we will comment on this in the discussion section.

So, as anticipated above, such a scalar needs a coupling to matter that is even weaker than that of spin 2. More precisely, the force between a pair of non-relativistic massive objects $M_1$ and $M_2$ is
\beq
{\bf F}_\phi=M_1\beta\,\nabla\phi_2=-{\beta^2 M_1 M_2\over 4\pi r^2}\hat{r}
\eeq
($\hat{r}$ is a radial unit vector). 
Compared to spin 2, we have $F_\phi/F_{GR}\approx\beta^2/(4\pi G_N)=2\beta^2\mpl^2$ (times $M_{\phi1}M_{\phi2}/M_1M_2$ if pressure is significant). So using the solar system bound on $\beta$, the new force is constrained to be 
\beq
{F_\phi\over F_{GR}}\leq1.25\times 10^{-5}
\eeq
Since the scalar does not couple to relativistic matter, this small ratio is essentially an {\em upper} bound on the ratio of forces. So for a canonical scalar it must remain essentially irrelevant to all physical processes since it must be at least 5 orders of magnitude weaker than known effects from spin 2. At first sight it would seem that this single observation of light bending eliminates any significant modifications to gravity, and one is stuck with the spin 2 theory which is general relativity.

\subsection{Making $\phi$ Relevant Again}

To avoid this conclusion, wherein $\phi$ is always irrelevant, one needs to make the scalar $\phi$ non-canonical. To do so, one needs to introduce non-trivial dynamics through the kinetic function $K(X)$ (or related higher derivative terms, such as in Galileon models). 

In much of the literature this involves kinetic terms that lead to {\em screening} on solar system scales, i.e., to a force that grows more slowly than $1/r^2$ in high density environments. 
On the one hand, this is quite an exciting idea, as it could potentially raise the bound on $\beta$ to be closer to $\sim 1/\mpl$. 
On the other hand, as we will discuss in the next subsection, this is unfortunately closely tied to superluminality. 

An alternative approach, which is the point of this paper, is to consider models in which the solar system bound on $\beta$ remains in place, but the kinetic functions $K(X)$ leads to a new force that can grow {\em faster} than $1/r^2$ in dense environments. This means that while the force can remain negligible on solar system scales, it can rise to large values, for example near neutron stars. Relatedly, the imposition of subluminality pushes us in this interesting direction, as we will show. This will be our new idea and focus in this work. 

We consider a class of models with kinetic functions $K(X)$ that enjoy a new scaling regime. Although the specific details are not necessarily important, a concrete class of models that we will examine is
\beq
K(X)=-{\T\over p}((1-X/\T)^p-1)
\label{KofX}\eeq
where $\T$ is a positive constant of units of energy density and $p$ is a constant exponent. For small values $|X|/\T\ll 1$, we recover the canonical value $K\approx X$, with low $X$ series expansion
\beq
K(X)=X+(1-p){X^2\over 2\T}+(1-p)(2-p){X^3\over6\T^2}+\ldots
\label{Kexpansion}\eeq
But for high density environments, with $X$ large and negative, $|X|/\T \gg 1$, we can enter a new regime in which
\beq
K(X)\approx-{\T\over p}(-X/\T)^p
\label{Kapprox}\eeq 
For reasons that will become clear in the next subsection, we will focus on the exponent $p$ in the range $1/2<p<1$. Note that at the edges of this range, we have quite familiar theories: for $p=1$ it is just a canonical scalar, for $p=1/2$ the kinetic term is precisely that of the Dirac-Born-Infeld (DBI) action for a scalar \cite{Silverstein:2003hf,Alishahiha:2004eh}. However, the intermediate values of $p$ lead to some interesting new physics. We note that in all regimes, including the large $|X|/\T\gg 1$ regime, one can show that these kinds of theories are rather stable against quantum corrections \cite{deRham:2014wfa}.

To see the new behavior, let us first write down the new equation of motion for $\phi$. In general it is
\beq
{1\over\sqrt{-g}}\,\partial_\mu\left(\sqrt{-g}\,K_X(X)\,g^{\mu\alpha}\partial_\alpha\phi\right) = -T_M{f'(\phi)\over 2\,f(\phi)}
\label{phifull}\eeq
where $K_X\equiv dK/dX$. 
In the weak field regime, we can again say $f'(\phi)\approx-2\beta$ and $g_{\mu\nu}\approx \eta_{\mu\nu}$. For a static point source, this becomes a kind of nonlinear version of the Poisson equation
\beq
-\nabla(K_X(X))\nabla\phi)=\beta\,T_M
\label{NonlinearPoisson}\eeq
with $X$ evaluated at $X=-(\nabla\phi)^2/2$. Note that $X$ can be large and negative; in such a regime our expression in Eq.~(\ref{KofX}) remains real valued even for fractional powers $p$.

\begin{figure}[t]
\centering
\includegraphics[width=12cm]{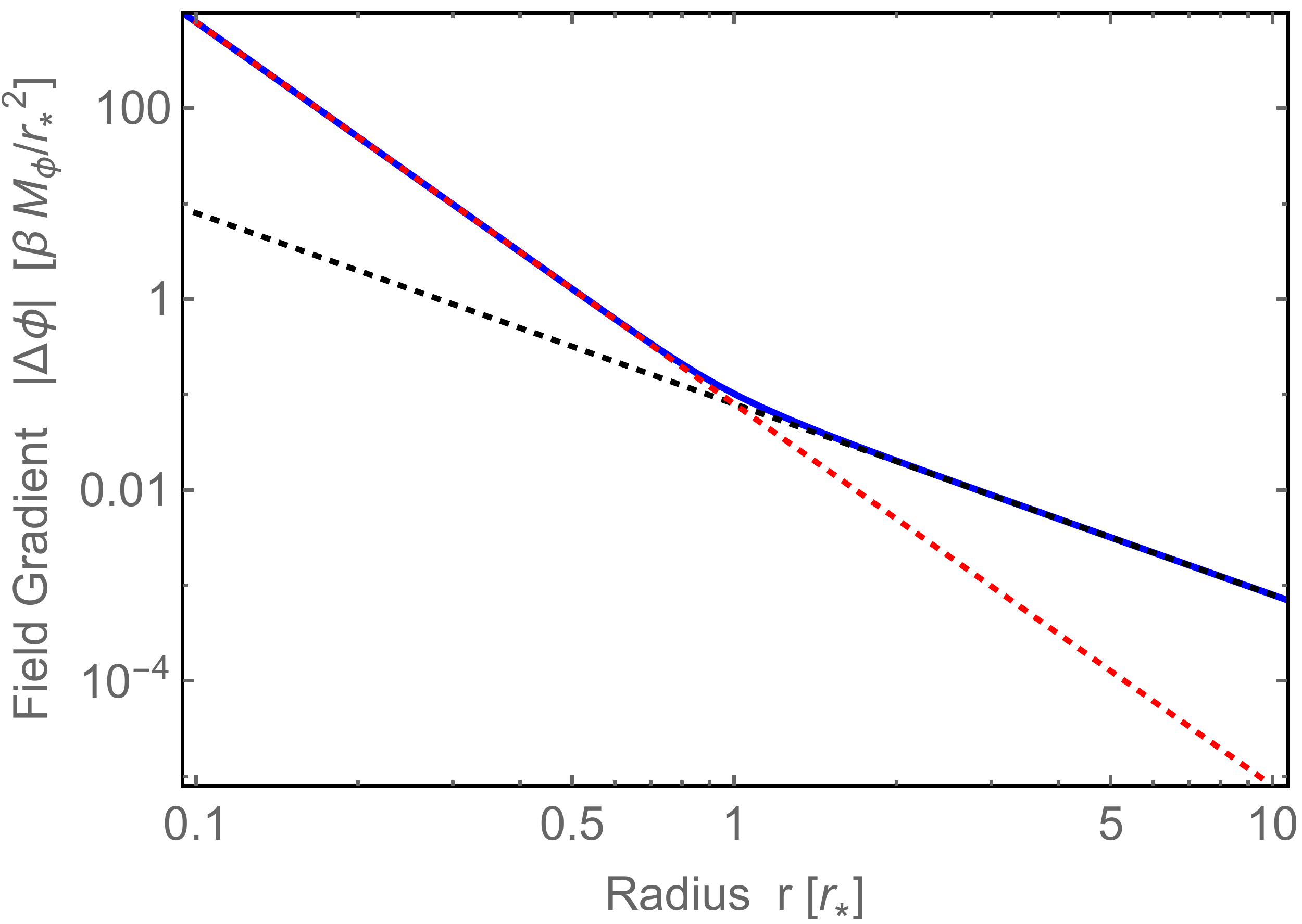} 
\caption{A plot of the field gradient $|\nabla\phi|$ (in units of $\beta M_\phi/r_*^2$) as a function of radius $r$ (in units of $r_*$) from solving Eqs.~(\ref{KofX},\,\ref{gradphisoln}). For illustration, we have chosen the exponent $p=3/4$ corresponding to $q=2$. The full curve is in solid blue. The limits of Eq.~(\ref{gradphilimits}) are also shown: Large $r\gg r_*$ standard regime is in dotted black. Small $r\ll r_*$ new regime is in dotted red.}
\label{fig:ForcePlot} 
\end{figure}

Consider a point source, with 
\beq
T_M=M_\phi\,\delta^3({\bf x})
\eeq
where $M_\phi =\int d^3x\,(\rho_M-3 \mathcal{P}_M)\approx \int d^3x\,(\rho-3 \mathcal{P})$ is typically close to the the total energy of the source $M$; but is more precisely the integrated trace of the energy-momentum tensor.  Then the formal solution to the modified Poisson equation is
\beq
K_X\left(-{(\nabla\phi)^2\over2}\right)\nabla\phi=-{\beta\,M_\phi\over 4\pi r^2}\,\hat{r}
\label{gradphisoln}\eeq
In principle one can then solve for $\nabla\phi$ depending on the choice of $K$. For the choice of $K$ given in Eq.~(\ref{KofX}) we can only do this numerically; see Figure \ref{fig:ForcePlot}. 
Importantly, we can derive limiting values analytically, as
\beq
\nabla\phi=-{\beta M_\phi\over 4\pi r^2}\,\hat{r}\times\Biggl\{\begin{array}{c}1,\,\,\,\,\,\,\,\,\,\,\,\,\,\,\,\,\,r\gg r_*\\
(r_*/r)^q,\,\,\,\,r\ll r_*
\end{array}
\label{gradphilimits}\eeq
where the exponent that is relevant in the near regime is
\beq
q={4(1-p)\over 2p-1}
\eeq
and the transition length scale we have defined is
\beq
r_*\equiv{\sqrt{\beta\,M_\phi}\over(2\T)^{1/4}\sqrt{4\pi}}
\eeq
This scale $r_*$ gives the characteristic boundary between two different scaling regimes. 
We see that for $1/2<p<1$, we have $q>0$ and so the scalar introduces a new force that is {\em steeper} than regular gravity in the $r\ll r_*$ regime. This is a kind of ``anti-screening" that we will focus on in this work (essentially the opposite of the Galileon or massive gravity models which ``screen" within the so-called Vainshtein radius). It means that even with $\beta\leq2.5\times 10^{-3}/\mpl$ to obey solar system bounds with a small $\propto 1/r^2$ force law at $r\gg r_*$, this new force can be relevant in very dense environments as it can transition to this new steep scaling force law $\propto 1/r^{2+q}$ at $r\ll r_*$. The ratio of forces on a test particle in this regime is approximately
\beq
{F_\phi\over F_{GR}} \approx 2\beta^2\mpl^2\left(r_*\over r\right)^{\!q}
\label{ForceRatio}\eeq
(times $M_{\phi1}M_{\phi2}/M_1M_2$ from pressure corrections). So this extra factor can make the relative effect of this new force quite important at small distances. 

We also note that this is quite unlike standard extra dimension models \cite{Arkani-Hamed:1998jmv}. For instance, if one introduces a fourth spatial dimension of appreciable size, it is well known this leads to a new $\propto 1/r^3$ force law. This is very tightly constrained from table top Cavendish experiments, requiring any extra dimensions to be extremely small and irrelevant in astrophysical environments. However, what we have here is a transition scale $r_*$ that is {\em environment dependent} (in this narrow sense, it is similar to the Vainshtein radius \cite{Vainshtein:1972sx}). In particular, while $\beta$ and $\T$ are fixed constants of the Lagrangian, the mass $M_\phi$ depends on environment. For light sources in table top Cavendish experiments, the mass $M_\phi$ is so small (perhaps a fraction of a kilogram) that the transition scale $r_*$ is incredibly tiny as it scales with mass $\sqrt{M_{\phi}}$. On the other hand, for very heavy sources, such as a neutron star with a huge mass $M_{\phi}\sim M_{sun}$, the transition scale can be astrophysically large, allowing new novel dynamics to take place. This will be our interest.

\subsection{Hyperbolicity and Subluminality Constraints}\label{HypSuperConstraints}

With the above considerations laid out, one should consider in what regime is the above Lagrangian well behaved theoretically. To this end we consider two serious theoretical issues: (i) ensuring hyperbolicity of the equations of motion and (ii) avoiding superluminality. In this paper, we shall refer to this as ensuring ``causality". Previous considerations of this topic include Refs.~\cite{Aharonov:1969vu,Arkani-Hamed,Bruneton:2006gf,Bruneton:2007si,Waldtextbook,Hertzberg:2021fvu,terHaar:2020xxb,Lara:2021piy}.

By expanding around a background configuration $\phi_0$ (see the analysis ahead in Section \ref{sec:Emission} for a related analysis) as $\phi=\phi_0+\pert$ one can show that small high energy perturbations $\pert$ obey the linear equation
\beq 
G^{\mu\nu}_\phi\partial_\mu\partial_\nu\pert=0
\eeq
where a kind of effective metric $G^{\mu\nu}_\phi$ is given by (here we follow our earlier work Ref.~\cite{Hertzberg:2021fvu})
\beq
G^{\mu\nu}_\phi=K_X(X_0)g^{\mu\nu}+K_{XX}(X_0)\partial^\mu\phi_0\partial^\nu\phi_0
\eeq
(with $K_X(X_0)\equiv dK/dX|_{X_0}$ and $K_{XX}(X_0)\equiv d^2K/dX^2|_{X_0}$ and $X_0=(\partial\phi_0)^2/2$ is the kinetic term evaluated on the background $\phi_0$).
In order to maintain hyperbolicity, we demand that the effective metric $G^{\mu\nu}_\phi$ carry the same signature of the underlying metric, which in our convention is $+---$. One can check that two eigenvalues of $G^{\mu\nu}_\phi$ are $-K_X$ which needs to be negative. While the determinant can be shown to be
\beq
\mbox{Det}[G^{\mu\nu}_\phi]=-K_X(X_0)^3(K_X(X_0)+2X_0K_{XX}(X_0))
\eeq
which also needs to be negative. So there are two conditions to maintain hyperbolicity, namely
\bea
&&(A)\,\,\,A\equiv K_X>0\\
&&(B)\,\,\,B\equiv K_X+2X K_{XX}>0
\eea
By inserting the above form for $K(X)$ into these expressions, we obtain
\bea
&&A=(1-X_0/\T)^{p-1}\\
&&B=(1-X_0/\T)^{p-2}(1-(2p-1)X_0/\T)
\eea
From the form of $K(X)$ it is clear that the physical phase space is the domain
\beq
X_0<\T
\label{Xps}\eeq
Note that since $X_0=(\dot\phi_0^2-(\nabla\phi_0)^2)/2$ we can have both positive or negative $X_0$, but the physical phase space bounds it from above. This guarantees that $A>0$ is always satisfied (in fact this justifies our choice of signs in the definition of $K$ to avoid negative energies). However, the sign of $B$ is less clear. We need the factor $(1-(2p-1)X_0/\T)>0$. If $1/2\leq p\leq 1$ then this is guaranteed by Eq.~(\ref{Xps}). If $p>1$, one can try restricting phase space further by demanding $X_0<\mu/(2p-1)$. 
However, if $p<1/2$ then we can not reasonably restrict phase space appropriately, as any reasonable theory must allow large negative values of $X_0$ (where the gradients are dominant; which are especially important near slowly varying sources) and $B<0$, violating hyperbolicity.

For hyperbolic equations, we can compute the speed of these high energy perturbations. One can check that the leading order perturbation from unity is
\beq
c_s^2 = 1\mp {2XK_{XX}\over K_X}
\eeq
where the $\mp$ depends on whether the background is primarily time dependent ($X_0>0$ and $-$) or primarily space dependent ($X_0<0$ and $+$). In either case, the condition to avoid superluminality is
\beq
(C)\,\,\,C\equiv K_{XX}\ge0
\eeq
By inserting the above for $K(X)$ into this expression, we obtain
\beq
C= (1-p)(1-X_0/\T)^{p-1}/\T
\eeq
Again using the phase space condition $X_0<\T$, we see that the system has superluminality if $p>1$, has luminality if $p=1$, and has subluminality if $p<1$. In fact for $p>1$, one can show that closed time-like curves can be constructed and so there is a definite form of acausality here \cite{Arkani-Hamed}.

Altogether, the conditions for a hyperbolic and non-superluminal theory are
\beq
1/2\leq p \leq 1
\eeq
Importantly, this implies that the $p>1$ regime, which leads to $q<0$ and a slower force law and screening, is associated with superluminality. While the $p<1/2$ regime, which does not possess spherically symmetric solutions at small radii anyhow, is associated with non-hyperbolicity. The $1/2<p<1$ regime, which is hyperbolic and causal, is precisely the regime that leads to ``anti-screening" with a force $\propto 1/r^{2+q}$ (with $q>0$), so that the scalar force can be quite important and interesting in very dense environments. 

As mentioned above, two familiar cases live at these boundaries: if $p=1$, then we have a canonical scalar, while if $p=1/2$, the above action is the kinetic part of the DBI action which can arise from the effective theory of a brane bending mode in extra dimension models. While neither of these cases is especially interesting to give rise to novel long-range force laws, it is curious that these two familiar cases are right on the edge of theoretical viability. 


\section{Corrections to Orbital Motion}\label{sec:Orbital}

\subsection{Motion in Binary System}

In this section we begin by computing the leading corrections to orbital motion of a binary system from the new scalar force. 
We will be utilizing the orbit equation in the following analysis. Let us give a quick derivation of it here. 

For a non-relativistic binary system with a two-body potential $V(r)$, we can use the standard one-body simplification by moving to center of mass co-ordinates. We define the reduced mass $\redu = M_1 M_2 / M$ and total mass $M = M_1 + M_2$. The planar equations of motion are
\bea
&&{dr\over dt} = \pm\sqrt{\frac{2}{\redu} \left( E- V(r) - \frac{L^2}{2\,\redu\, r^2} \right)}\\
&&{d\theta\over dt} = \frac{L}{\redu\, r^2}\label{thetadoteqn}
\eea
where $r$ is radius and $\theta$ is angle in the plane, $E$ is the conserved energy, and $L$ is the conserved angular momentum.
By dividing these equations to eliminate $dt$, we obtain the following differential relation
\beq
d\theta = \frac{L}{\redu\, r^2} \frac{1}{\sqrt{\frac{2}{\redu} \left( E - V(r) - \frac{L^2}{2\redu r^2} \right)}} dr 
\eeq
If we now perform the standard substitution $u \equiv 1/r$, we can represent the angular change in integral form
\beq
\Delta \theta = - \int_{u_0}^u \frac{du}{\sqrt{\frac{2\redu E}{L^2} - \frac{2\redu V(u)}{L^2} - u^2}} 
\label{DeltaThetaV}\eeq

To make progress we need an explicit form for the potential $V(r)$. As a starting point, let us recap the standard case of general relativity, which in this limit is just the Newtonian form $V(r) = - \frac{A}{r}$, where $A = G \redu M$. In terms of $u$, this is $V(u) = - A \,u$. Inserting this, the integral is simple. Re-arranging gives the famous elliptic orbit
\beq
r(\theta) = \frac{a(1 - \epsilon^2)}{1 + \epsilon \cos(\theta - \theta_0)} 
\eeq
where the eccentricity $\epsilon$ and semi-major axis $a$ are given by
\bea
&&\epsilon = \sqrt{1 + \frac{2EL^2}{\redu A^2}} \\
&&a=-{A\over 2E}
\eea
(note that the non-relativistic energy $E$ is negative in a bound orbit, so that $\epsilon<1$ and $a$ is positive).

\subsection{Orbital Motion for $p=5/6$}


We now wish to compute the correction to the motion from the new scalar force. In the linear regime the scalar force is also $\propto 1/r^2$ and so this is immediately degenerate with the spin 2 force in the non-relativistic regime. More precisely, it just adjusts the effective Newton constant from the bare value in the Lagrangian to the observed $G$. Of course there are then residual effects when studying the motion of ultra-relativistic motion, such as light, but this is not the focus of this section. Instead, if we pass to the nonlinear regime $r\ll r_*$, we obtain a new force $\propto 1/r^{2+q}$ which is no longer degenerate. 

Although an analysis for general $q$ is interesting, we shall illustrate the idea here with the case of $q=1$, which corresponds to $p=5/6$. For this case, in the nonlinear regime the scalar force on a test particle $M_1$ from a source $M_2$ is of the form
\beq
{\bf F}_\phi = -{\beta^2M_1M_2\,r_{*2}\over 4\pi r^3}\hat{r}
\eeq
where $r_{*2}=\sqrt{\beta\,M_2}/((2\T)^{1/4}\sqrt{4\pi})$ refers to the transition scale from source $M_2$. However, we in fact need to insert a two-body potential energy function $V(r)$ into the non-relativistic theory. This must be symmetric between particles 1 and 2. Since we don't precisely have superposition in a nonlinear theory, the full potential energy function is nontrivial between objects of comparable masses. For simplicity, we can approximate the pairwise energy function as 
\beq
V(r) = -{G\redu M\over r}\left(1+{\Bs\over 2r}\right)
\eeq
where we have defined a useful length scale associated with the scalar 
\beq
\Bs=2\beta^2\mpl^2\,r_{*}
\label{Forcep56}\eeq
(times $M_{\phi1}M_{\phi2}/M_1M_2$ if pressure of sources is significant). 
We will take $r_*=\mbox{Max}\{r_{*1},r_{*2}\}$. When either $M_1$ or $M_2$ is much larger than the other, this is accurate. However, when $M_1\sim M_2$ there may be an $\mathcal{O}(1)$ error in this estimate.
Although we will enter the nonlinear $r\ll r_*$ regime, the smallness of $\beta^2\mpl^2$ means that this new term can still be treated perturbatively with $\Bs\ll r$.

Let's follow the methodology outlined above, and begin by finding the orbit equation. We insert this potential into Eq.~(\ref{DeltaThetaV}) to obtain the integral
\beq
\Delta \theta = - \int_{u_0}^u \frac{du}{\sqrt{\frac{2\redu E}{L^2} + \frac{2\redu A}{L^2}u - \Big(1 - \frac{\redu A\Bs}{L^2} \Big) u^2}} 
\eeq
This shows the convenience of the $q=1$ choice that we are making for illustrative purposes; the term in the denominator inside the square root is still only a quadratic function of $u$. Therefore it can still be integrated in closed form. Doing so and inverting gives
\beq
u = \frac{1}{r(\theta)} = \frac{A\, \redu}{L^2 - A\Bs \redu} \Bigg( 1 + \sqrt{1 + \frac{2 L^2 E}{\redu A^2} - \frac{2E}{A} \Bs} \  \cos\Big( \sqrt{1 - \frac{\redu}{L^2} A\Bs} \  \theta \Big) \Bigg) 
\eeq
This is again an elliptic orbit; due to our special choice of $p=5/6$. However, one sees that the argument of the cosine and the amplitude are altered.

\subsection{Tensor Power Output and Rate of Change of Period}

Since the orbit is corrected, one might wonder if there is a corresponding alteration in the tensor power output.
The (time-averaged) power lost into tensor modes $P_{t}$ has a well known form in terms of derivatives of the quadrupole moment tensor $Q_{ij}$
\beq
P_t =- \frac{G}{5} \Bigg{\langle}\left( \dddot{Q}_{ij} \dddot{Q}^{ij} - \frac{1}{3} \dddot{Q}^2  \right) \Bigg{\rangle}
\eeq
where $Q$ is the trace of $Q_{ij}$. For an elliptical orbit in the $xy$-plane, the quadrupole moment takes the form
\beq
Q_{ij} = \redu \, r(\theta)^2 
\begin{pmatrix}
\cos^2\theta & \sin\theta \cos\theta & 0\\
\sin\theta \cos\theta & \sin^2\theta & 0\\
0&0&0
\end{pmatrix}
\eeq
and we have used angled brackets $\langle\ldots\rangle$ to indicate a time average.

Now, with $r=r(\theta)$ a given orbit, we have $Q_{ij}$ as purely a function of $\theta$, which is quite useful. We need one more tool before we can proceed; We notice that if directly compute $\dddot{Q}_{ij}$, the final answer will involve several high-order derivatives of $\theta$. In order to remedy this, we can express $\dot{\theta}$ using Eq.~(\ref{thetadoteqn}).
Using this, we can at each derivative step replace $\dot{\theta}$ with $L/(\redu\,r(\theta)^2)$, which is an explicit function of $\theta$. This leaves us with no derivative terms in the $\dddot{Q}_{ij}$ expression. With this, we can find $\dddot{Q}_{ij}$ and in turn the power.

Carrying out this algebra and working to leading order in $\Bs$ we find that this leads to the result
\beq
P_t=P_{GR}\left(1+\gamma\,{2\beta^2\mpl^2\,r_*\over a}\right)
\label{PGRcorrection}\eeq
where $P_{GR}$ is the standard Peter's power formula from general relativity of
\beq
P_{GR}=-{32G^4M_1^2M_2^2(M_1+M_2)(1+73\epsilon^2/24+37\epsilon^4/96)\over 5 a^5(1-\epsilon^2)^{7/2}}
\eeq
and the dimensionless coefficient of our correction is
\beq
\gamma = {12(24+104\epsilon^2+33\epsilon^4)\over(1-\epsilon^2)(96+292\epsilon^2+37\epsilon^4)}
\eeq
We see that typically the relative correction to the power $\delta P_t/P_t\approx 2\gamma\,\beta^2\mpl^2\,r_*/a$ is on the order of the correction to the force $\delta F/F\approx F_\phi/F_{GR}\approx 2\beta^2\mpl^2\,r_*/a$, with $r$ evaluated at a typical point on the trajectory, i.e., on order the semi-major axis $a$. If the trajectory is highly elliptical, with $1-\epsilon^2$ very small, then there is a further enhancement from $\gamma$.






The orbit is no longer periodic in the presence of the scalar force. In this case, the usual definition of its period that is relevant to observation is to define the period $\Per$ as the time it takes from perihelion to perihelion. Using the above equations, we can write this as
\beq
\Per={\redu\over L}\int_0^{\theta^*}d\theta\,r(\theta)^2
\eeq
where a completed orbit is no longer $\theta^*=2\pi$, but it is now
\beq
\theta^*={2\pi\over\sqrt{1-{\redu\over L^2}A\Bs}}
\eeq
By carrying out the integral we find that the dependence on $\Bs$ drops out, and we are left with the canonical result
\beq
\Per={\pi A\sqrt{\redu}\over\sqrt{2}\,(-E)^{3/2}}
\eeq

As the system emits power, it loses energy and its parameters change adiabatically slowly. We can use this to compute the rate of change of period as
\bea
{d\Per\over dt}={d\Per\over dE}{dE\over dt}
={3\pi A\sqrt{\redu}\over 2\sqrt{2}(-E)^{5/2}}\,P_t
\eea
By re-writing the semi-major axis as $a=-A/(2E)$ and re-writing energy in terms of period $\Per$, the rate of change of period can be expressed as
\beq
{d\Per\over dt}={d\Per\over dt}\Bigg{|}_{GR}\left(1+\gamma{2\beta^2\mpl^2\,r_*\over a(\Per)}\right)
\eeq
with $a(\Per)=(G(M_1+M_2)\Per^2/4\pi^2)^{1/3}$ is the Kepler law, which is unchanged in this special case.

Let us mention that for the Hulse-Taylor binary system \cite{Hulse:1974eb}, the measured value of the intrinsic period change (after accounting for galactic corrections) and the value predicted by general relativity is \cite{Weisberg:2016jye}
\beq
{{d\Per\over dt}\Big{|}_{obs}\over{d\Per\over dt}\Big{|}_{GR}}=0.9983\pm 0.0016
\eeq
So this shows agreement at the $\sim 10^{-3}$ level; which is the famous success that general relativity predicts the period shortening of the Hulse-Taylor system accurately due to tensor mode emission. Hence our new scalar force must not alter this too much. This bounds our correction to be no more that $\sim 10^{-3}$. Using the eccentricity of $\epsilon_{\mbox{\tiny{HT}}}\approx 0.617$, we obtain $\gamma\approx 6.2$. Hence on the scale of the Hulse-Taylor system, which has a semi-major axis of $a_{\mbox{\tiny{HT}}}=1,950,000$\,km, the new force should obey
\beq
{F_\phi\over F_{GR}}\lesssim 2\times10^{-4}\,\,\,\,\,\,\,\,\mbox{on scale $a_{\mbox{\tiny{HT}}}=1,950,000$\,km}
\eeq
Recall that the solar system's linear bound is $F_\phi/F_{GR}\leq 1.25\times 10^{-5}$. These moderately larger values can arise as we transition into the nonlinear regime as assumed here. However, as we shall now demonstrate in the next subsection, there is a much more serious constraint from precession.

\subsection{Precession}

As mentioned above, in one orbit of angle $\theta\to2\pi$, we see that the radius does not return to its original value. Hence the orbit is not periodic and we have precession, even in this non-relativistic limit. In this non-relativistic regime, the correction arises entirely from this new force. The shift in the semi-major axis per orbit is
\beq
\Delta\sigma_\phi= {2\pi\over\sqrt{1 - \frac{\redu}{L^2} A\Bs}}-2\pi\approx
{\pi\,\redu\,A\,\Bs\over L^2}
\eeq
where we have expanded to leading order in $\Bs$ in the second step.  To first approximation, the angular momentum in a Keplerian orbit is given by
\beq
L\approx\sqrt{A\,a\,\redu(1-\epsilon^2)}
\eeq
Inserting this into the above gives a leading estimate of the precession
\beq
\Delta\sigma_\phi={\pi\,B_\phi\over a(1-\epsilon^2)} = {2\pi\beta^2\mpl^2\,r_*\over a(1-\epsilon^2)}
\label{precessionPhi}\eeq

Let us compare this to the famous result for the precession from the leading relativistic effect in general relativity
\beq
\Delta\sigma_{GR}
= {6\pi G M\over a(1-\epsilon^2)}
\eeq
If we now form the ratio of these we obtain
\beq
{\Delta\sigma_\phi\over \Delta\sigma_{GR}}={\beta^2\mpl^2\,r_*\over 3 GM}
\eeq
We note that this is potentially quite significant. Recall that $\beta^2\mpl^2 r_*/a$ is on the order of the ratio of forces. 
While the typically orbital speed is $v_{\mbox{\tiny{orb}}}\sim\sqrt{GM/a}$. Hence we can write this as roughly
\beq
{\Delta\sigma_\phi\over \Delta\sigma_{GR}}\sim{F_\phi\over F_{GR}}\,{1\over v_{\mbox{\tiny{orb}}}^2}
\label{PrecessionRatio}\eeq
Hence the effect is {\em enhanced} relative to the ratio of forces. 

As shown in the very interesting work Ref.~\cite{Barreira:2015aea} (which focussed on superluminal models), in the case of a roughly circular orbit (ignoring corrections from $\epsilon$) one can readily obtain a general formula for the precession for any kinetic function $K$ as
\beq
\Delta\sigma_\phi \approx -{8\pi\beta^2\mpl^2\,X\,K_{XX}\over K_X(K_X+2XK_{XX})}
\label{PrecessionGeneral}\eeq
One can check that in the $p=5/6$ theory and deep in the nonlinear regime, this recovers our result above (\ref{precessionPhi}) when ignoring ellipticity. More generally (see ahead to Eq.~(\ref{KXexp})) in this highly nonlinear, but roughly circular limit, this becomes
\beq
\Delta\sigma_{\phi}\approx{8\pi(1-p)\beta^2\mpl^2\over 2p-1}\left(r_*\over a\right)^q
\eeq
Curiously, in the regime of interest $1/2<p<1$, this is always positive, as in general relativity. However the magnitude can be very different. 
By comparing to general relativity, and recalling the force ratio (\ref{ForceRatio}), this once again exhibits the scaling mentioned above in Eq.~(\ref{PrecessionRatio}). 

In general this means that if we are deeply in the nonlinear regime these models give rise to potentially very large corrections to the precession. This is because the effect from the scalar occurs even in the non-relativistic limit, while the general relativistic result is precisely due to relativistic corrections, and is $\sim v_{\mbox{\tiny{orb}}}^2$ suppressed. So for example, if the moon or mercury or the Hulse-Taylor pulsar system is in the nonlinear regime, this put a severe constraint on the model by demanding that the ratio of forces is extremely small to compensate for (i) the relative factor of $1/v_{\mbox{\tiny{orb}}}^2$ and (ii) the fact that the precession is well measured in these systems and in good agreement with general relativity. One can of course proceed by demanding $\beta$ is orders of magnitude smaller than the solar system bound. But a perhaps more interesting way to proceed is to ensure that the transition scale $r_*$ is smaller than the orbital radius in these systems so they are in the linear regime. By demanding $a\gg r_*$ (where $a$ is the semi-major axis of the orbit) we obtain the corresponding bound on the cutoff of the theory of:
\bea
&&\T^{1/4}\gg 0.2\,\mbox{eV}\left(\beta\over 2.5\times 10^{-3}/\mpl\right)^{1/2}\,\,\,\,\,\mbox{(from moon;} \,\,a_{\mbox{\tiny{moon}}}=3.84\times 10^5\,\mbox{km})\\
&&\T^{1/4}\gg 0.9\,\mbox{eV}\left(\beta\over 2.5\times 10^{-3}/\mpl\right)^{1/2}\,\,\,\,\,\mbox{(from mercury;} \,\,a_{\mbox{\tiny{mercury}}}=5.79\times 10^7\,\mbox{km})\\
&&\T^{1/4}\gg 30\,\mbox{eV}\left(\beta\over 2.5\times 10^{-3}/\mpl\right)^{1/2}\,\,\,\,\,\,\mbox{(from Hulse-Taylor;} \,\,a_{\mbox{\tiny{HT}}}=1.95\times 10^6\,\mbox{km})
\eea
Of these cases, the Hulse-Taylor system appears to place the most restrictive constraint, which is reasonable since it involves extremely heavy objects in rather close proximity. On the other hand, a direct comparison to theory in the precession of Hulse-Taylor is not easy, since the measured precession is in fact used to {\em infer} the masses of the neutron stars, i.e., there is degeneracy. Conversely, the moon and mercury have independent measurements to resolve degeneracies and so have a well defined and successful prediction for the precession within general relativity.

\subsection{Quasi-Linear Regime}

Let us suppose that indeed we are not in fact in the fully nonlinear regime in a well measured binary system (let's assume $\mu^{1/4}>30$\,eV, for example), but instead we are in the quasi-linear regime. This means that we can to first approximation use the linear scalar force theory. In the non-relativistic limit this does not give rise to any precession as it also gives $\propto 1/r^2$ force, much like Newtonian gravity. It does lead to a very small precession when relativistic effects are taken into account, but since $\beta^2\mpl^2<10^{-5}$, this is even much smaller than the general relativity prediction, and so can be ignored.

However, even if a system is deeply in the linear regime, i.e., $r\gg r_*$, as we will ensure mercury or the moon is, there is still some residual nonlinearity in the theory which can give rise to some precession. We would like to compute that here. 

Let's consider general values of $p$. For $r\gg r_*$ the modified Poisson equation (\ref{NonlinearPoisson}) exhibits an asymptotic series expansion, which we find to be
\beq
\nabla\phi=-{\beta\,M_\phi\over 4\pi r^2}\,\hat{r}\left(1+(1-p)\left(r_*\over r\right)^{\!4}+\ldots\right)
\eeq
It is simple to see why the leading correction is $\propto 1/r^4$ relative to the first; this is because the leading approximation of $\nabla\phi\propto 1/r^2$ gives $X\propto 1/r^4$ to leading order, and hence the leading correction to $K(X)$ is $1/r^4$ suppressed; see Eq.~(\ref{Kexpansion}). 
The leading term here does not lead to precession, but the second term does. By inserting this into the above formula for the precession Eq.~(\ref{PrecessionGeneral}), we obtain the leading order result
\beq
\Delta\sigma_{\phi}\approx 8\pi(1-p)\,\beta^2\mpl^2\left(r_*\over a\right)^{\!4}
\eeq
By noting that in this nearly linear regime, the ratio of forces is simply $F_\phi/F_{GR}\approx 2\beta^2\mpl^2$, then the ratio of precession has the scaling
\beq
{\Delta\sigma_\phi\over\Delta\sigma_{GR}}\sim{F_\phi\over F_{GR}}\left(r_*\over a\right)^{\!4}{1\over v_{\mbox{\tiny{orb}}}^2}
\eeq
This is much more reasonable. In this regime, the ratio of forces is already known to be bounded by $F_\phi/F_{GR}\leq 1.25\times 10^{-5}$, as we discussed in Section \ref{SSB}. Moreover, the large $1/v_{\mbox{\tiny{orb}}}^2$ factor can be compensated by $(r_*/a)^4$ factor, which is self-consistently very small in this nearly linear regime. 

In the next section where we study stability of neutron stars, we will show that we have a bound on the cutoff of $\T^{1/4}\gtrsim 200\,\mbox{eV}(\beta\,\mpl/(2.5\times 10^{-5}))^{5/2}$ for $p=5/6$. Using this as a convenient reference, we can express the scalar precession for the moon, mercury, and Hulse-Taylor as
\bea
&& \Delta\sigma_\phi\approx 4.6\times 10^{-17}\left(1-p\over 1/6\right)\left((200\,\mbox{eV})^4\over\T\right)\left(\beta\over 2.5\times 10^{-5}/\mpl\right)^4\,\,\,\,\,\mbox{(moon)}\\
&& \Delta\sigma_\phi\approx 1.0\times 10^{-14}\left(1-p\over 1/6\right)\left((200\,\mbox{eV})^4\over\T\right)\left(\beta\over 2.5\times 10^{-5}/\mpl\right)^4\,\,\,\,\,\mbox{(mercury)}\\
&& \Delta\sigma_\phi\sim 10^{-8}\left(1-p\over 1/6\right)\left((200\,\mbox{eV})^4\over\T\right)\left(\beta\over 2.5\times 10^{-5}/\mpl\right)^4\,\,\,\,\,\,\,\,\,\,\,\,
\,\,\,\,\,\,\,\,\,\,\mbox{(Hulse-Taylor)}
\eea
where in the Hulse-Taylor case we have only included an estimate, since the masses of the two neutron stars are comparable and there is significant ellipticity, so our treatment is not quite precise in this regime.

The precession of the moon as it orbits the earth, according to general relativity is $\Delta\sigma_{GR}\approx  2\times 10^{-10}$, which is nearly 7 orders of magnitude larger than the above representative value. Since it is measured with accuracy of only a few significant figures, this is easily compatible.
The precession of mercury is well known to be beautifully compatible with general relativity's prediction of $43\,\mbox{arc-sec/century}$; converted into radians per orbit, this is $\Delta\sigma_{GR}\approx  5\times 10^{-7}$. This is nearly 8 orders of magnitude larger than the above value and hence this is also compatible. 

Finally, the Hulse-Taylor binary has a measured precession of \cite{Weisberg:2016jye}
\beq
\Delta\sigma_{obs}\approx6.52342\times 10^{-5}(1\pm 10^{-6})\,\,\,\,\,\,\,\,\,\,\,\,
\,\,\,\,\,\,\,\,\,\,\mbox{(Hulse-Taylor)}
\eeq
The central value here is only 3 to 4 orders of magnitude larger than the above value. While the uncertainty is fantastically small at the $\sim 10^{-10}$ level, which is in fact smaller than the above value $\Delta\sigma_\phi$. However, the prediction from general relativity is slightly unclear here, because the precession is in fact used to infer the masses in the binary. 

What we can say is that we should presumably not alter the inferred masses appreciably, or else the prediction for the power emitted in tensor modes would be in error. The agreement between the general relativity prediction for the power output (inferred through the period change) is within an accuracy of $\sim 10^{-3}$, as we mentioned at the end of the last subsection. So a simple interpretation is that we should not alter the precession by a (relative) factor of $10^{-3}$. This brings the observed value of $6.5\times 10^{-5}$ down to $6.5\times 10^{-8}$, which is approaching the above reference value for the precession. Taking these parameters, we obtain the bound
\beq
\T^{1/4}\gtrsim 125\,\mbox{eV}\left(|1-p|\over 1/6\right)^{\!1/4}\left(\beta\over 2.5\times 10^{-5}/\mpl\right)
\,\,\,\,\,\,\,\,
\,\,\,\,\,\,\,\,\,\,\mbox{(Hulse-Taylor)}
\label{PrecessionConst}\eeq
Other competitive bounds arise from measurements of the double pulsar (PSR J0737-3039A/B) \cite{Burgay:2003jj}, which is moderately more compact than the Hulse-Taylor system, though more circular, and whose precession is measured slightly less accurately. In any case, we shall use Hulse-Taylor as a useful guide.

\begin{figure}[t!]
\centering
\includegraphics[width=12cm]{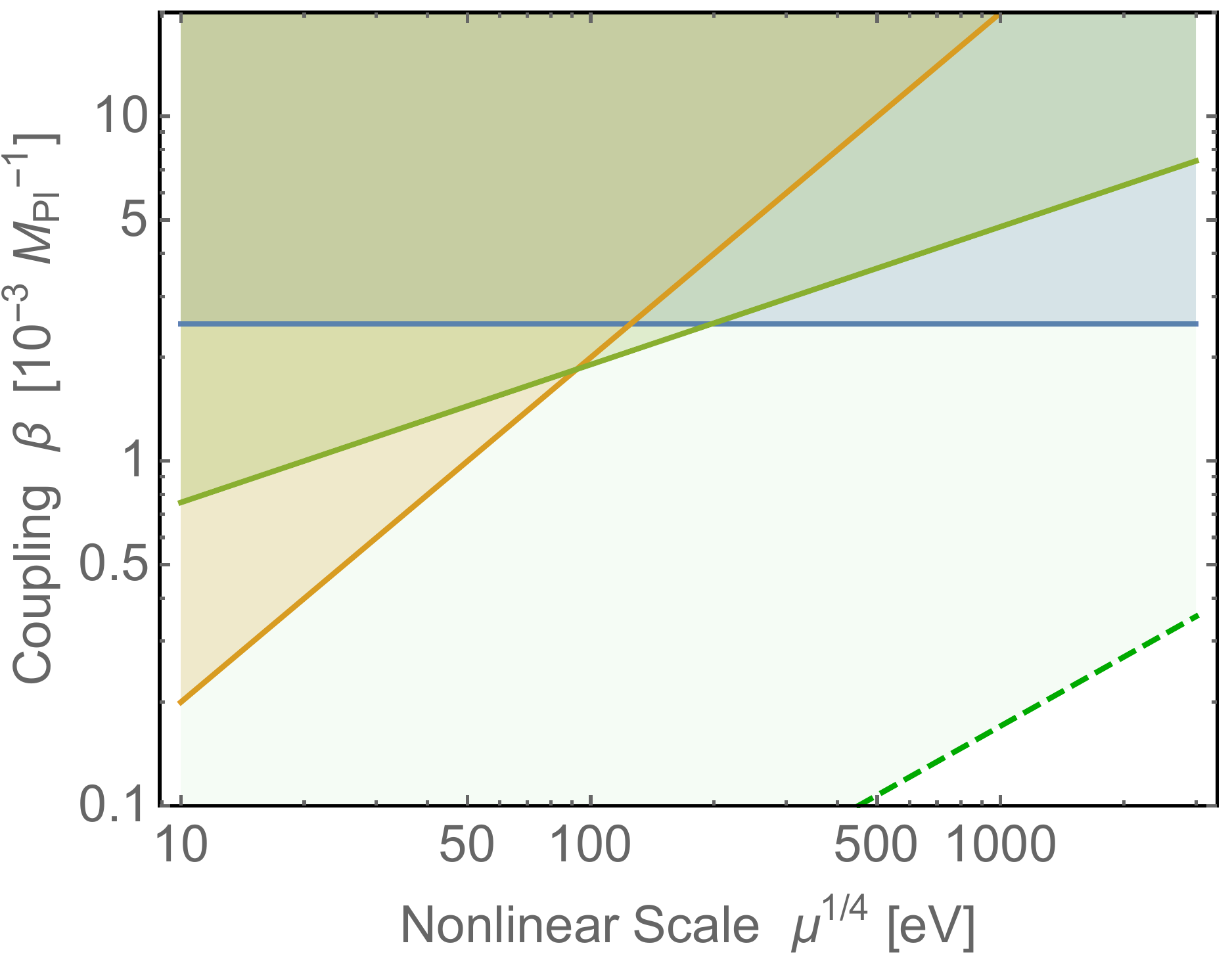} 
\caption{Observational constraints on the model parameters: Coupling $\beta$ on vertical axis and nonlinear scale $\T^{1/4}$ on horizontal axis. The blue region is ruled out by solar system tests of light bending. The orange region is ruled out by precession of Hulse-Taylor pulsar binary; this assumes $p\sim 5/6$, but it is rather insensitive to its value (assuming $p$ is not extremely close to $1$). The green region is ruled out by altering neutron stars significantly and is sensitive to the value of $p$; the solid green line is for $p=5/6$, the dashed green line is for $p=3/4$. In general, for lower $p$ the green region becomes larger, for higher $p$ the green region becomes smaller. The remaining lower right region is allowed (depending on $p$) according to our analysis here (albeit other phenomena, such as mergers, could affect this).}
\label{fig:ConstraintPlot} 
\end{figure}

We will obtain a complementary bound in the next Section. 
Combining the standard constraint from solar system tests in Eq.~(\ref{betaconstraint}), with our new constraint from Hulse-Taylor precession in Eq.~(\ref{PrecessionConst}) and the upcoming neutron star equilibrium constraint in Eq.~(\ref{mubound}) or (\ref{mubound2}), we obtain the plot in Figure \ref{fig:ConstraintPlot}. The colored regions are ruled out, while the blank (lower right) region is allowed. There can be further constraints from mergers, as we discuss later. 

\section{Hydrostatic Equilibrium of Stars}\label{sec:Equilibrium}

As we saw in the previous section, the constraint from Hulse-Taylor binary is that the new force must remain considerably smaller than the standard gravitational force or the orbital motion will be highly altered. In fact it must be the quasi linear regime, or the precession is altered too much. Now even if this is obeyed, we should recognize that new scalar force will still enter the nonlinear regime for dynamics on much smaller scales scales. For the case $p=5/6$, this then gives a $1/r^3$ as opposed to $1/r^2$ force law, which is an appreciably steep rise. So by considering much closer material, the force can rise to be quite large.

A very important issue to address is what happens at the surface or interior of neutron stars themselves. Since they have a radius of $\sim 10$\,km, this reduction in $r$ by about $10^5$ compared to Hulse-Taylor orbit is very large indeed. For the $p=5/6$ and $1/r^3$ force law, this could lead to a new force comparable to, or larger than, regular gravity. So in this section, we can use the existence and stability of neutron stars to impose corresponding bounds on the parameters $\T$ and $\beta$. This is important for $p=5/6$ or for even smaller $p$ values, where the rise in forces is even steeper. On the other hand, this new analysis may not be so relevant for $p$ that is only slightly less than 1, say $p=11/12$, giving rise to a $1/r^{2.4}$ force law, which is only a mild enhancement relative to $1/r^2$ when we move in towards the surface of the neutron star $\sim 10$\,km from the Hulse-Taylor orbit $\sim 10^{6}$\,km.

\subsection{Static and Spherically Symmetric}

In the vicinity of a neutron star, we need the full equations of general relativity, as we are approaching the strong field regime, along with the full equations for the scalar $\phi$. We will model a single neutron star as spherically symmetric and in equilibrium it is static in its reference frame. To this end we can decompose the metric into a standard form that exploits its spherically symmetry as
\beq
ds^2=g_{\mu\nu}dx^\mu dx^\nu = e^{2\alpha(r)}dt^2-e^{2\gamma(r)}dr^2-r^2d\theta^2-r^2\sin^2\theta \,d\varphi^2
\eeq
where we have introduce two functions of radius $\alpha(r)$ and $\gamma(r)$. 

The Einstein field equations take on the standard form, albeit we have a more complicated source than usual. We write it as
\beq
G_{\mu\nu}=8\pi G\left(T^{(M)}_{\mu\nu}+K_X(X)\partial_\mu\phi\partial_\nu\phi- g_{\mu\nu}K(X)\right)
\eeq
where the first term on the right hand side is the energy-momentum tensor of the matter sector (computed with the conformal coupling $f(\phi)$) and the remaining terms are the direct contributions to the energy-momentum tensor from the scalar's kinetic term. 

To make progress, we need to assume a form for the matter sector's $T^{(M)}_{\mu\nu}$. For simplicity, we shall assume that the matter is described by a perfect fluid, with energy density $\rho$ and pressure $\mathcal{P}$. In the absence of the conformal coupling to $\phi$, this gives rise to the energy-momentum tensor
\beq
\hat{T}^{(M)}_{\mu\nu} = -\mathcal{P}\,g_{\mu\nu}+(\rho+\mathcal{P})U_\mu U_\nu
\eeq
where the hat indicates that this is defined {\em without} the coupling to $\phi$. Here $U_\mu$ is the (covariant) 4-velocity. For a static configuration, the only non-zero component is $U_t=e^\alpha$ to ensure normalization ($g^{\mu\nu}U_\mu U_\nu=1$). Hence
\beq
\hat{T}^{(M)}_{\mu\nu}=\mbox{diag}\left[\rho\,e^{2\alpha},\mathcal{P}\,e^{2\gamma},\mathcal{P}\,r^2,\mathcal{P}\,r^2\sin^2\theta\right]
\eeq

Now let us re-instate the dependence on the conformal coupling to $\phi$. For a collection of massive point particles, primarily carrying only gravitational interactions, their action is
\beq
S_{M}=\sum_i-m_i\int d\lambda\sqrt{g_{\mu\nu}{dx_i^\mu\over d\lambda}{dx_i^\nu\over d\lambda}}\sqrt{f(\phi)}
\eeq
where we have used the fact that they are coupled via $\tilde{g}_{\mu\nu}=f(\phi)g_{\mu\nu}$. Carrying through the procedure to obtain the energy momentum tensor, we see that the final result in this case just factorizes as
\beq
T^{(M)}_{\mu\nu}= \hat{T}^{(M)}_{\mu\nu}\sqrt{f(\phi)}
\eeq

\subsection{Modified TOV Equation}

We now insert the above into the Einstein field equations. For the $tt$-equation, it is convenient to trade in the metric function $\gamma(r)$ for the enclosed mass $m(r)$ through
\beq
e^{2\gamma(r)}=\left(1-{2Gm(r)\over r}\right)^{-1}
\label{gammaeqn}\eeq
Then one finds that the $tt$-Einstein equation becomes
\beq
{dm\over dr}=4\pi\,r^2\rhoeff(r)
\label{masseqn}\eeq
where the total effective energy density is
\beq
\rhoeff=\sqrt{f(\phi)}\,\rho-K(X)
\label{rhoeffeqn}\eeq
with $X$ taking on the value
\beq
X=-{1\over2}\left(1-{2Gm(r)\over r}\right)(\phi')^2
\label{Xeqn}\eeq
where $\phi'\equiv d\phi/dr$. 
One can prove that by integrating over all space, then $m(r\to\infty)=M$ is the conserved total energy of the system. 

The $rr$-equation gives us a differential equation for $\alpha$ of
\beq
{d\alpha\over dr}={G(m(r)+4\pi r^3 \peff)\over r(r-2Gm(r))}
\label{alphaeqn}\eeq
where the total effective pressure is
\beq
\peff=\sqrt{f(\phi)}\,\mathcal{P}+K(X)+\left(1-{2Gm(r)\over r}\right)K_X(X)(\phi')^2
\label{peffeqn}\eeq
Then by using using the covariant conservation of energy momentum $\nabla^\mu T_{\mu\nu,\mbox{\tiny{eff}}}=0$ with $\nu=r$, one obtains 
$(\rhoeff+\peff)d\alpha/dr=-d\peff/dr$. 
So altogether we obtain a kind of updated version of the Tolman-Oppenheimer-Volkoff (TOV) equation \cite{Tolman:1939jz,Oppenheimer:1939ne}, where we simply need to replace $\rho$ and $\mathcal{P}$ by $\rhoeff$ and $\peff$, to obtain
\beq
{d\peff\over dr}=-{G(\rhoeff+\peff)(m(r)+4\pi r^3\peff)\over r(r-2Gm(r))}
\label{peqn}\eeq

Also, we need the differential equation for $\phi$. We find this to be the following second order ODE
\beq
{1\over e^{\alpha(r)+\gamma(r)}\,r^2}\,{d\over dr}\!\left(e^{\alpha(r)-\gamma(r)}r^2 K_X(X){d\phi\over dr}\right)=(\rho-3\mathcal{P}){f'(\phi)\over 2\sqrt{f(\phi)}}
\label{phieqn}\eeq
($f'\equiv df/d\phi$) which can be further expanded out by using Eqs.~(\ref{gammaeqn}) and (\ref{alphaeqn}) to eliminate $\gamma(r)$ and $\alpha(r)$ from this equation completely.

Finally, we need an equation of state for our matter, relating pressure $\mathcal{P}$ to density $\rho$. We shall be interested in neutron stars. As a first approximation (e.g., see Ref.~\cite{WeinbergGravitation}), we shall model this as a system of free degenerate fermions. In reality there are significant corrections to this from QCD, but this is a useful starting point. It is well known that for free fermions at degeneracy the energy density and pressure can be expressed as an integral over wavenumber, with cutoff at the Fermi wavenumber $k_F$, as
\bea
&&\rho = 2\int_0^{k_F} \!{d^3k\over(2\pi)^3}\,\sqrt{k^2+m_n^2}\label{rhofermi}\\
&&\mathcal{P} = {2\over3}\int_0^{k_F} \!{d^3k\over(2\pi)^3}\,{k^2\over\sqrt{k^2+m_n^2}}\label{pfermi}
\eea
where 
$m_n$ is the neutron mass. 
This provides a parametric relationship between pressure and energy density, through the Fermi wavenumber $k_F$. At low and high densities, we have the limiting cases
\bea
&&\mathcal{P}={\rho^{5/3}\over5\,\rho_n^{2/3}},\,\,\,\,\rho \ll \rho_n\\
&&\mathcal{P}={\rho\over3},\,\,\,\,\,\,\,\,\,\,\,\,\,\,\rho\gg \rho_n
\eea
where we have introduced a characteristic neutron star energy density of
\beq
\rho_n\equiv {m_n^4\over 3\pi^2}
\eeq

So our system of equations is (\ref{masseqn},\,\ref{peqn},\,\ref{phieqn}), along with the definitions (\ref{rhoeffeqn},\,\ref{Xeqn},\ref{peffeqn}) and equation of state given implicitly through Eqs.~(\ref{rhofermi},\,\ref{pfermi}). In addition we need a choice of kinetic function $K(X)$, which we take to be of the form specified earlier in Eq.~(\ref{KofX}). And we need a choice of conformal coupling function $f(\phi)$, which we take to be
\beq
f(\phi)=(1-\beta\,\phi)^2
\eeq
which has the nice property that the right hand side of Eq.~(\ref{phieqn}) is just $-\beta(\rho-3\mathcal{P})$.

We also need boundary conditions to complete the system. We impose $m(0)=0$, so the total enclosed mass of zero radius is itself zero. We impose $\phi'(0)=0$, so that the scalar's differential equation is well behaved at the origin. Finally, for a localized source, we impose $\phi(r\to\infty)=0$, since we are defining $\phi=0$ as our asymptotic vacuum. We these boundary conditions imposed, we then have only one free parameter, which we take to be the energy density at the center 
\beq
\rho_c\equiv\rho(0)
\eeq
which parameterizes a family of star solutions.

\subsection{Numerical Results}

\begin{figure}[t!]
\centering
\includegraphics[width=10cm]{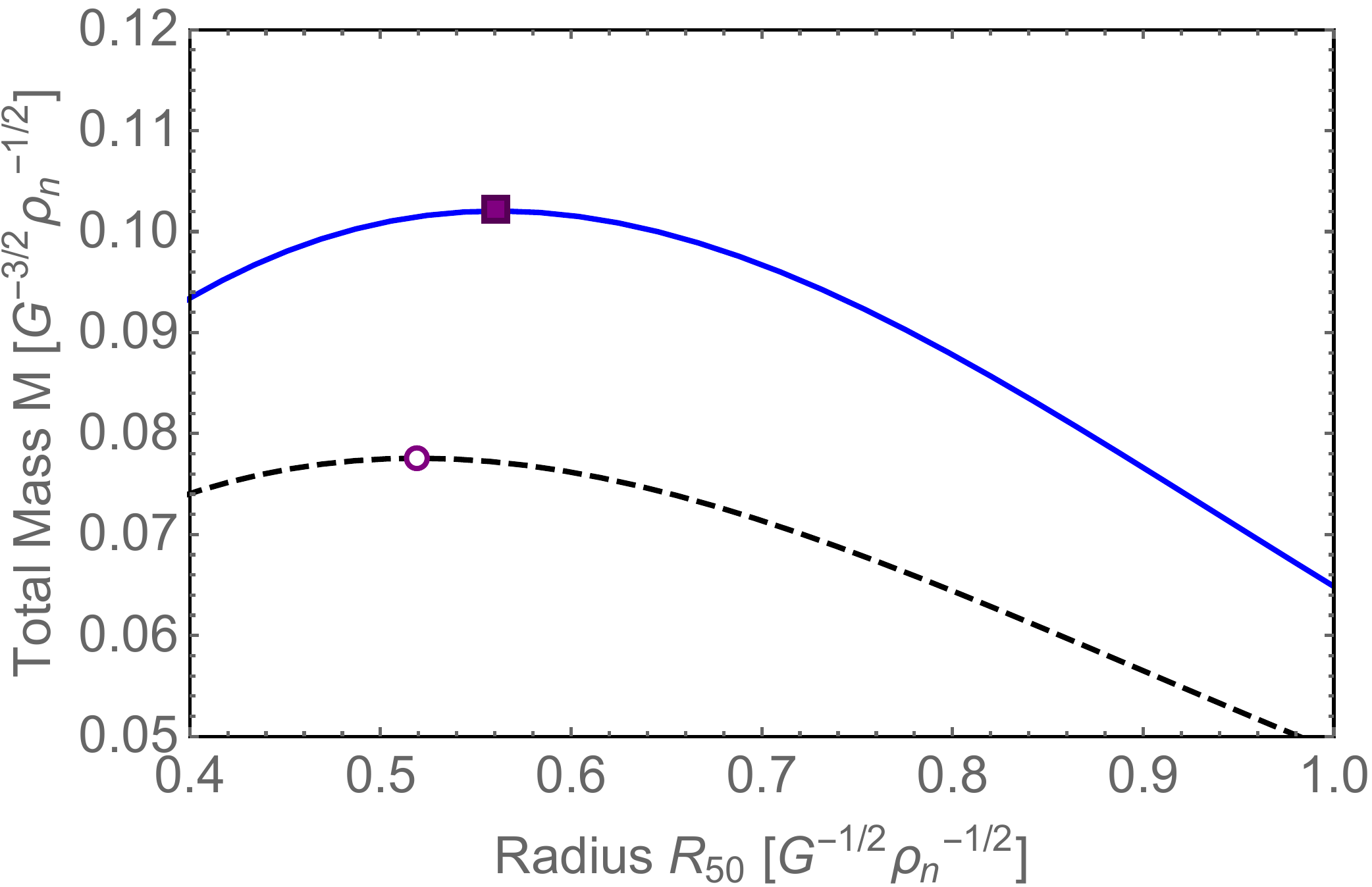}
\caption{The total integrated mass of the star versus the radius at which the density drops by a factor of 50. The lower black dashed curve is for pure general relativity. The upper solid blue curve includes our coupling to the scalar $\phi$ with $\beta=2.5\times 10^{-3}/\mpl$, $\T^{1/4}=7.4\times 10^{-7}\,\rho_n^{1/4}$, $p=5/6$. The regime of stability is when the derivative is negative (right side of purple marker). The regime of instability is when the derivative is positive (left side of purple marker).}
\label{fig:FigMassRadius} 
\end{figure}

\begin{figure}[t!]
\centering
\includegraphics[width=10cm]{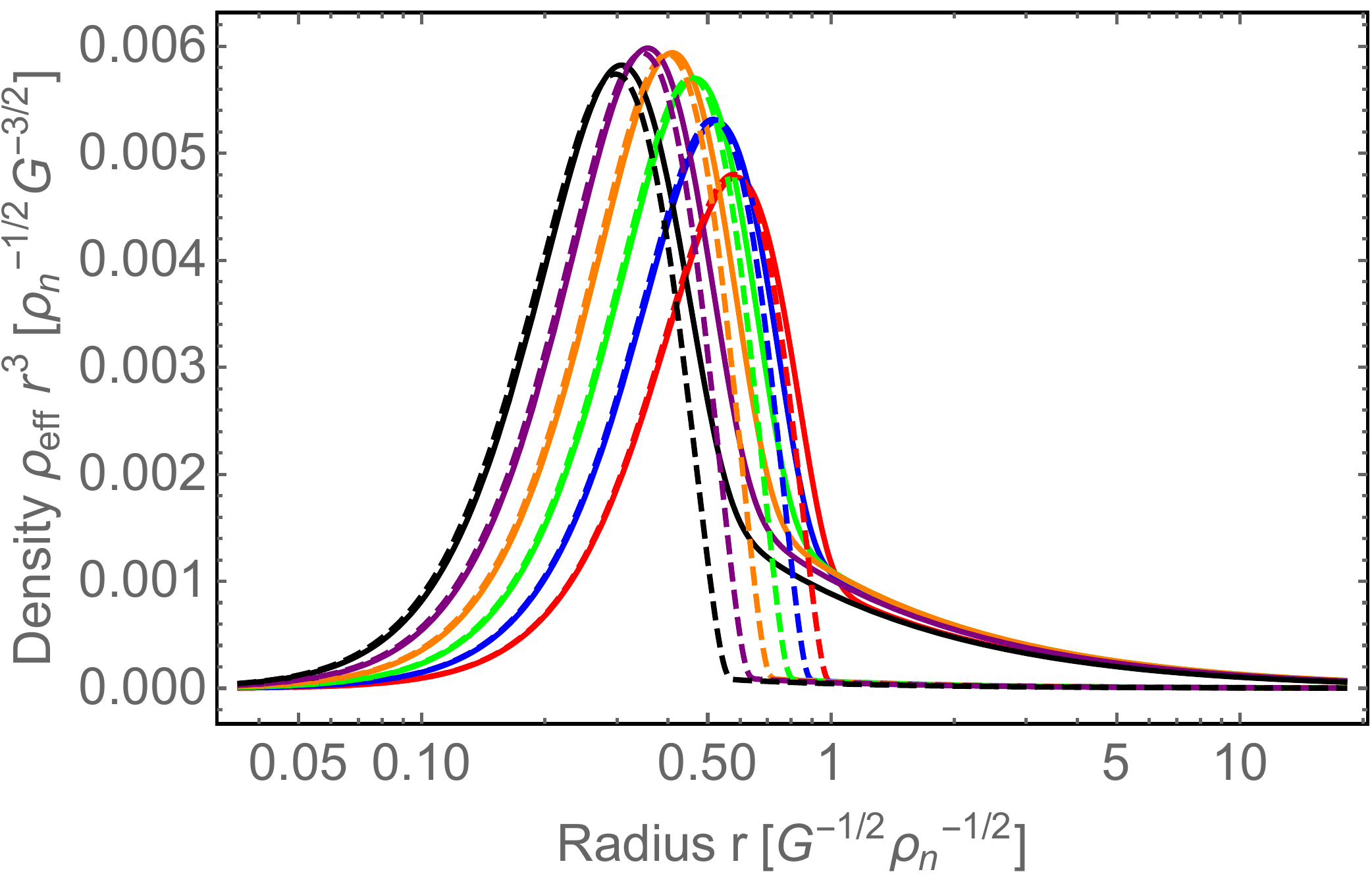}

\vspace{0.8cm}

\includegraphics[width=10cm]{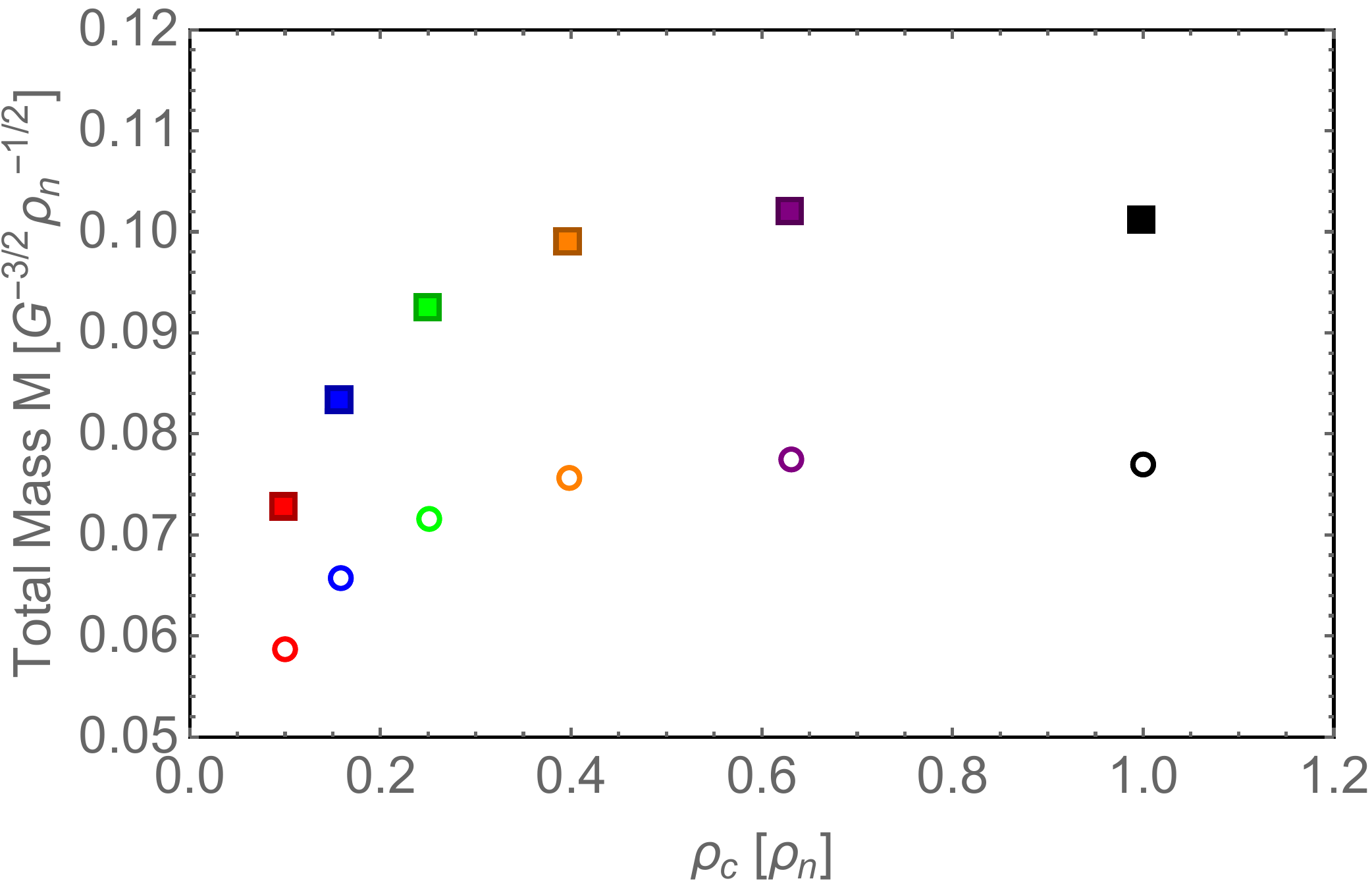} 
\caption{Top panel: Total energy density $\rhoeff$ (re-scaled by $r^3$ for convenience) in a neutron star as a function of radius. The pure general relativity result is given by the dashed lines. The result including the coupling to the scalar $\phi$ with $\beta=2.5\times 10^{-3}/\mpl$, $\T^{1/4}=7.4\times 10^{-7}\,\rho_n^{1/4}$, $p=5/6$ is given by the solid lines. The central density $\rho_c$ increases from red ($0.1\,\rho_n$) to blue ($0.16\,\rho_n$) to green ($0.25\,\rho_n$) to orange ($0.40\,\rho_n$) to purple ($0.63\,\rho_n$) to black ($1.0\,\rho_n$) as indicated by the bottom panel. Bottom panel: The total integrated mass of the star. The lower open circles are for pure general relativity. The upper filled squares include our coupling to the scalar. The increase in mass in the presence of the scalar is related to the extended radius of the star due to additional pressure from the scalar field (see Fig.~\ref{fig:FigForceRatio}).}
\label{fig:FigDensityrescaled} 
\end{figure}

\begin{figure}[t!]
\centering
\includegraphics[width=10cm]{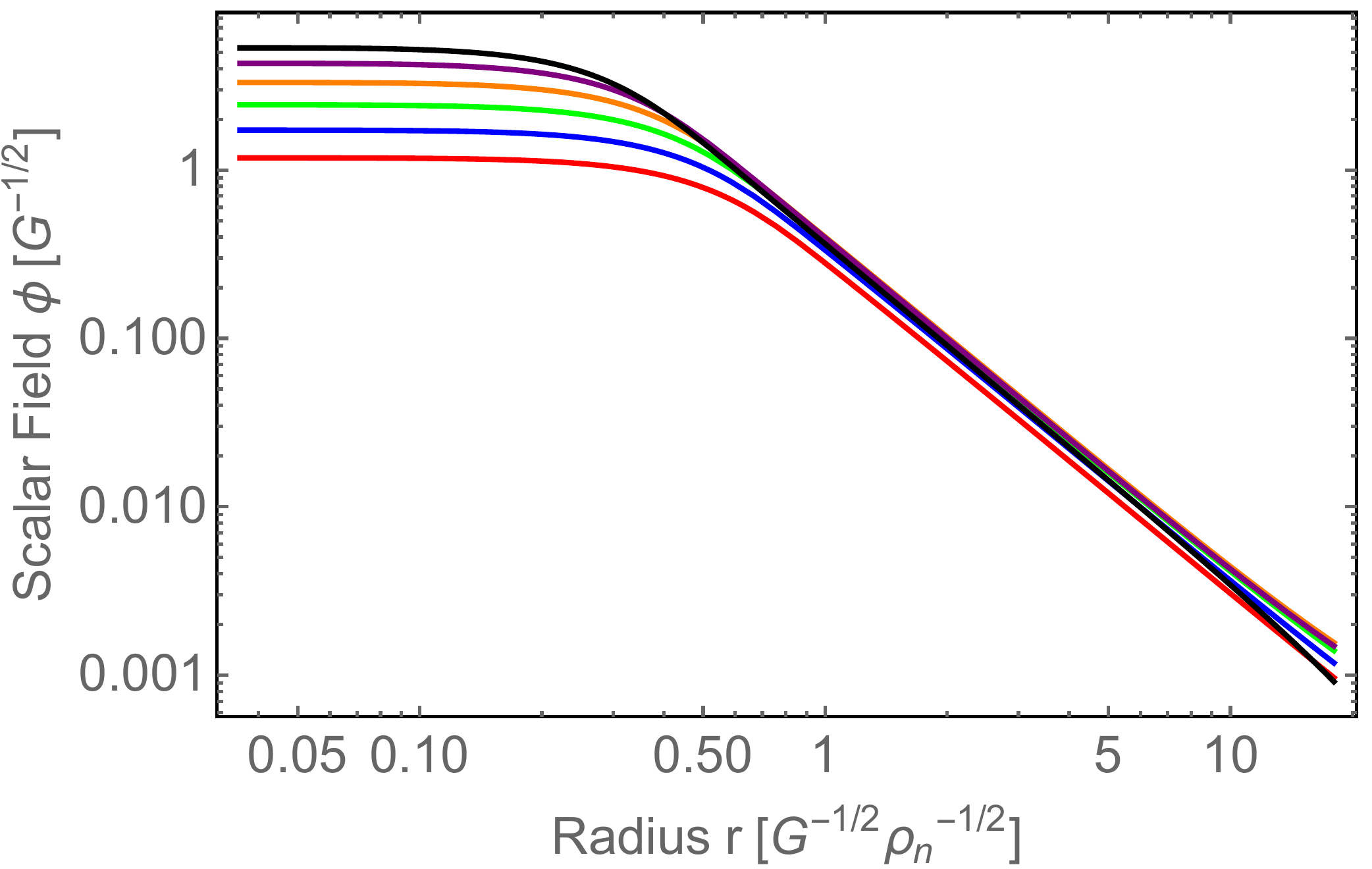} 

\vspace{0.8cm}

\includegraphics[width=10cm]{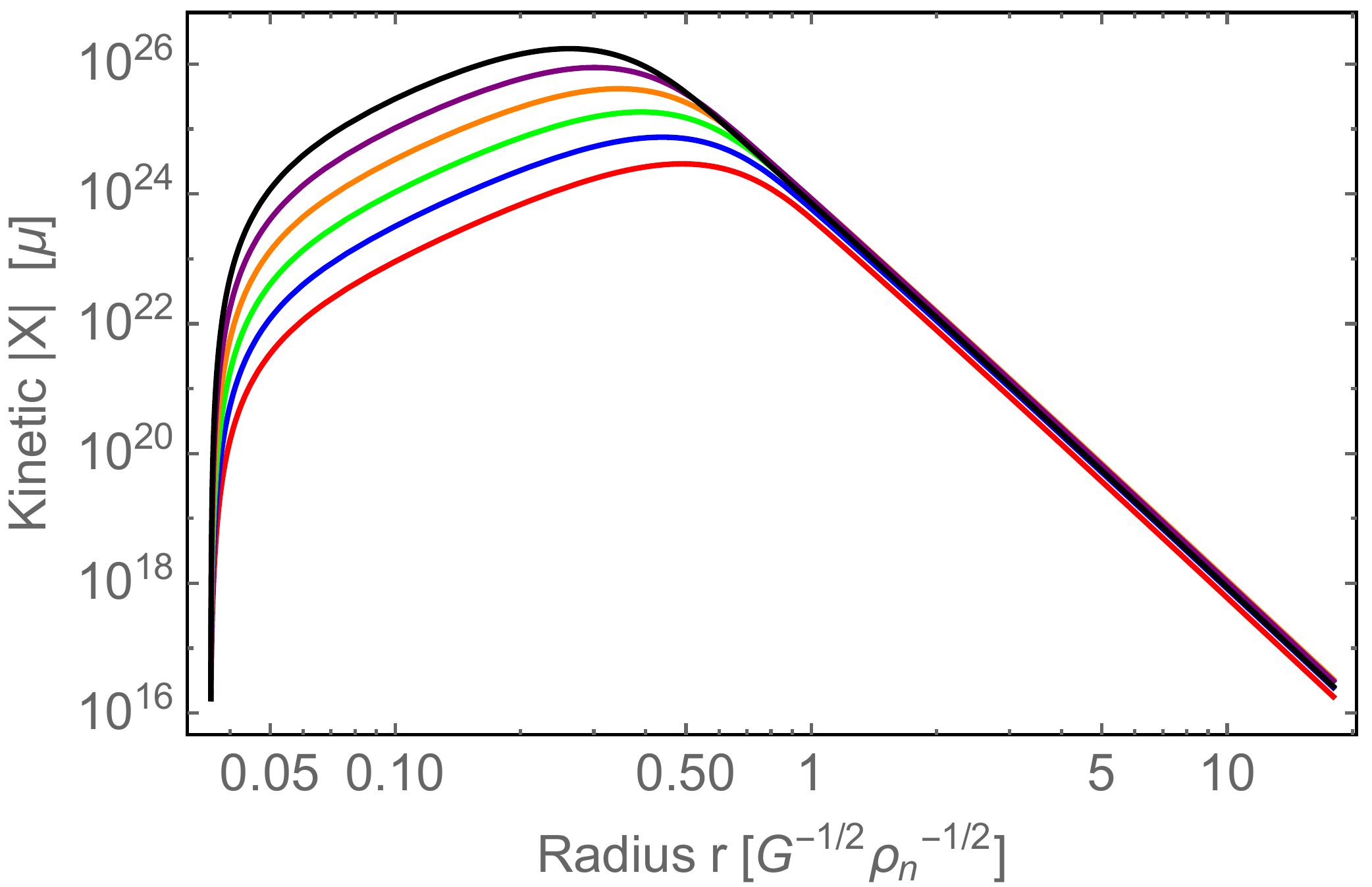} 
\caption{Top panel: The scalar field $\phi$ as a function of radius in a neutron star. Bottom panel: The corresponding kinetic term $|X|$ in units of $\T$. The parameters of the model and the color choices are the same as in Fig.~\ref{fig:FigDensityrescaled}.}
\label{fig:FigScalarField} 
\end{figure}

\begin{figure}[t!]
\centering
\includegraphics[width=10cm]{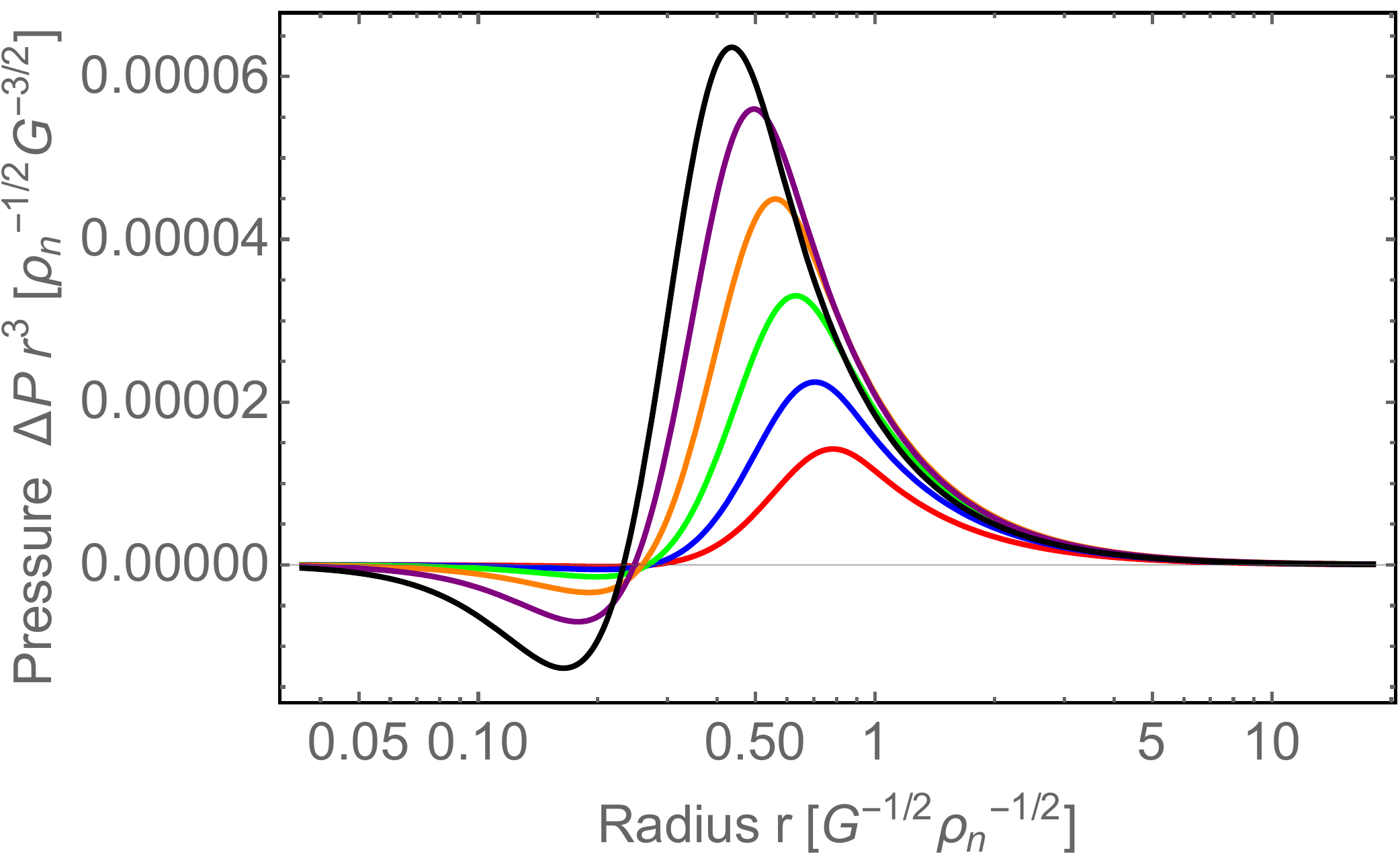} 

\vspace{0.8cm}

\includegraphics[width=10cm]{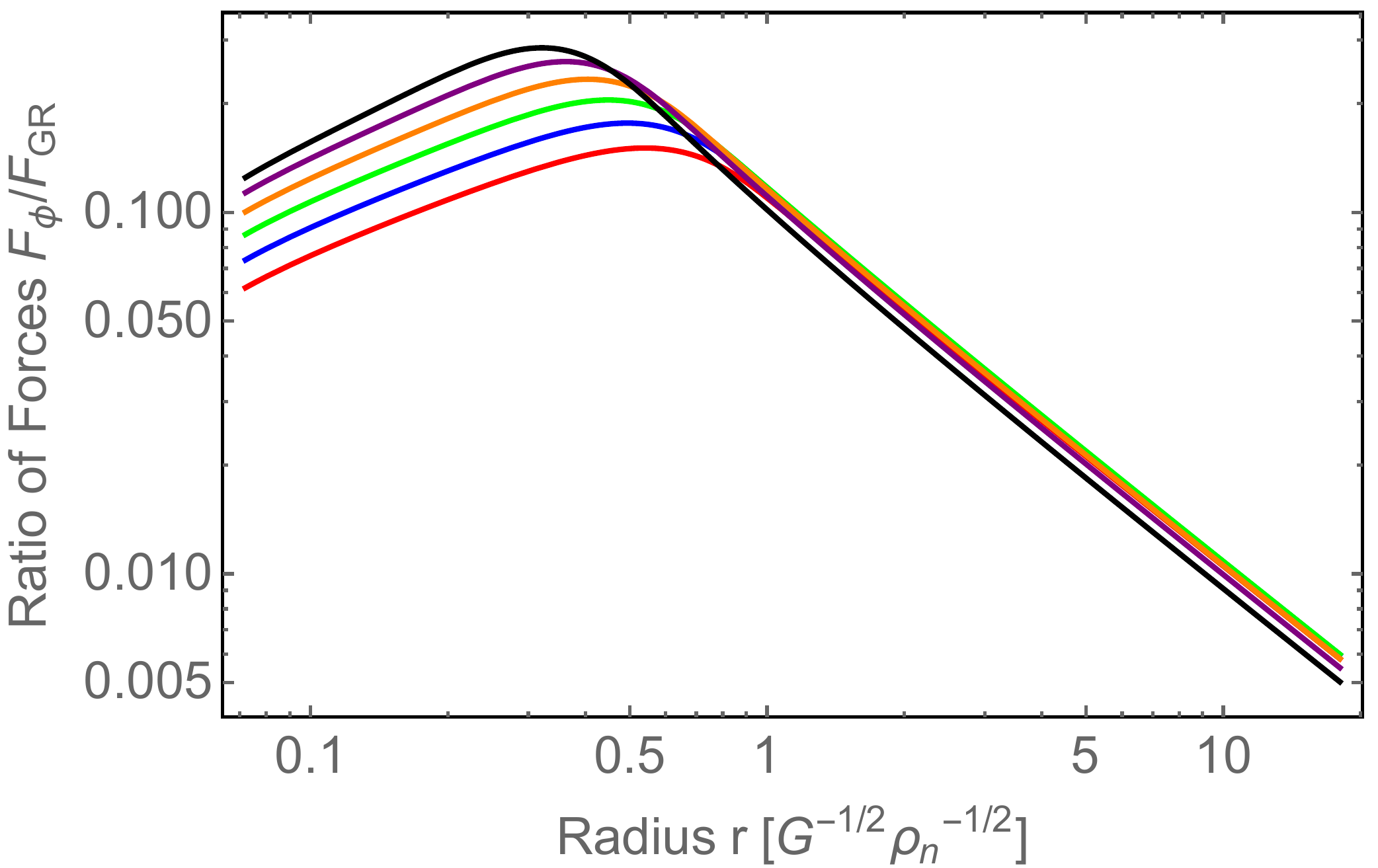} 
\caption{Top panel: The difference $\Delta \mathcal{P}\equiv\peff-\mathcal{P}$ between the effective pressure that sources gravitation and just the neutron pressure; we see that the scalar field creates a considerable amount of positive pressure which explains the extended radius of the star seen in Fig.~\ref{fig:FigDensityrescaled}).
Bottom panel: The ratio of forces (accelerations) on a unit test mass in a neutron star of the scalar $F_\phi=\beta|\nabla\phi|$ and standard gravity in the Newtonian approximation $F_{GR}=Gm(r)/r^2$. 
The parameters of the model and the color choices are the same as in Fig.~\ref{fig:FigDensityrescaled}. Note that outside the neutron star, the ratio is decreasing, since $F_\phi\propto 1/r^3$ and $F_{GR}\propto 1/r^2$. However at very large radii $r\gg r_*$, the ratio is constant, but that region is not visible here.}
\label{fig:FigForceRatio} 
\end{figure}

We have numerically solved the above system of differential equations. For $\beta=2.5\times 10^{-3}/\mpl$, $\T^{1/4}=7.4\times 10^{-7}\,\rho_n^{1/4}$, and $p=5/6$ the results are given in Figures \ref{fig:FigMassRadius},\,\ref{fig:FigDensityrescaled},\,\ref{fig:FigScalarField},\,\ref{fig:FigForceRatio}.  

In Fig.~\ref{fig:FigMassRadius} we plot the total integrated mass versus radius of the star. 
In this model, the radius is not quite as sharply defined as in standard gravity, as the scalar field provides a pressure that makes the density decrease towards zero less sharply (see below for discussion). For a concrete definition, we have taken the radius to be the value at which the density drops from its central value by a factor of 50; we have denoted this $R_{50}$.

In Fig.~\ref{fig:FigDensityrescaled} we show the energy density $\rhoeff$ versus radius, comparing the standard general relativity result (dashed lines) to the new result including the coupling to the scalar (solid lines). The radius of the star is increased compared to the standard result. This occurs because there is significant pressure stored in the scalar field; see ahead to Fig.~\ref{fig:FigForceRatio}. This increased pressure can therefore support even more mass from gravitational collapse. Correspondingly, the total energy is significantly higher, as seen in the bottom panel.

For small to moderate central core densities $\rho_c$, the stars are stable. However, beyond a certain central density, there is an instability.
This occurs for approximately the same central density range as in standard general relativity. In the mass-radius figure \ref{fig:FigMassRadius}, the instability occurs when the derivative $dM/dR$ changes sign to be positive at small radii (e.g., see Ref.~\cite{WeinbergGravitation}). This is at the purple point in the general relativity case (circle) and the new model case (square).
So we see that the minimum radius for stability (purple point) is shifted to moderately larger radii than in general relativity. So instead of the minimum radius of $\sim10$\,km, it may be $15$\,km, or so, depending on parameters $\beta$ and $\mu$. Enhanced radii are therefore a feature of this class of models.

In Fig.~\ref{fig:FigScalarField} we show the scalar field profile versus radius. Towards the center of star it is roughly flat. However, outside the star it falls off as a power law. We have checked that outside the star it indeed matches that expected from a point source, with asymptotic behavior given in Eq.~(\ref{gradphilimits}). In the regime plotted, we are only in the nonlinear regime $r\ll r_*$, so here it scales as $\propto 1/r^3$ for $p=5/6$ ($q=1$). The lower panel of this figure confirms that we are deeply in the nonlinear regime as $|X|\gg \T$ here. Of course at large $r\gg r_*$ it will transition to the linear regime.

In Fig.~\ref{fig:FigForceRatio} we plot the ratio of the new scalar force to the spin 2 force. At the center of star, the ratio approaches zero, since the center is highly relativistic and scalars do not couple to relativistic matter. Outside the star, we see the ratio decrease, since the scalar force is $\propto 1/r^3$ (for $p=5/6$; $q=1$), while the spin 2 force is $\propto 1/r^2$. At very large radii they will both scale as $1/r^2$, but this regime is not displayed. As the plots shows, the biggest correction from the scalar force to the neutron star is therefore in the outer layers of the neutron star. For the parameters chosen in the plot, the ratio of forces reaches an upper value of $\approx 0.3$.

We can anticipate that if the ratio of forces is increases significantly further, we will alter neutron stars appreciably. In fact we already see from Fig.~\ref{fig:FigDensityrescaled} that in the presence of significant coupling to scalars, the mass of the neutron stars is appreciably raised. Since current observations are roughly compatible with general relativity, this indicates we should impose a bound on this new force. However, since individual neutron stars are not fully understood, this bound is only rough.
For neutron stars (with $p=5/6$ or $p=3/4$), we can impose that the changes to the neutron star are no more than a factor of 2  (the factor of 2 is illustrative; if we move this value moderately, then there is only a small change in the constraint region), this provides the lower bound on our parameter of
\bea
\T^{1/4}\gtrsim 200\,\mbox{eV}\left(\beta\over 2.5\times 10^{-3}/\mpl\right)^{5/2}\,\,\,\,\,\,(\mbox{for}\,\,\,\,p=5/6;\,\,q=1)\label{mubound}\\
\T^{1/4}\gtrsim 60\,\mbox{keV}\left(\beta\over 2.5\times 10^{-3}/\mpl\right)^{3/2}\,\,\,\,\,\,(\mbox{for}\,\,\,\,p=3/4;\,\,q=2)
\label{mubound2}
\eea
Note that this bound should not be too surprising. If we considered the case $\beta\sim 1/\mpl$, then this bound says that $\T^{1/4}$ is constrained to be at least on the order of a GeV, which is roughly the neutron mass. This naturally says that the cross over energy density to the nonlinear scale $\T$ must be at least the neutron star density. However, for smaller $\beta$ to accommodate solar system constraints, the constraint on $\T$ scales accordingly. 

Interestingly, for $p=5/6$ when $\T$ is at this boundary with $\beta=2.5\times 10^{-3}/\mpl$, the cross over length scale $r_*$ is on the order of $r_*\sim 300,000$\,km, which is less than half the solar radius. This means that at the solar radius the scalar force has essentially transitioned into the linear regime, and so standard solar system constraints should indeed be obeyed. If, on the other hand, we lower $\beta$ to be even smaller than $2.5\times 10^{-3}/\mpl$, while simultaneously still ensuring $\T$ obey the neutron star bound Eq.~(\ref{mubound}), then the nonlinear regime is held fixed, while the linear regime has a suppressed force and so solar system tests are even more easily obeyed. 
Furthermore, for the Hulse-Taylor pulsar system, it has a closest approach (periastron) of $\approx 750,000$\,km, and semi-major axis of $\approx 1,950,000$\,km, so it is indeed in the nearly linear regime as required from the previous Section. Also the double pulsar (PSR J0737-3039A/B) system has a semi-major axis $\approx 900,000$\,km and is quite circular; so it is in the nearly linear regime too.

For higher $p$ values, the constraint on $\mu$ becomes weaker, as the rise in force is more moderate. Conversely, for lower $p$ values, the constraint on $\mu$ becomes stronger, as the rise in force is more significant. At the boundary, the transition scale $r_*$ is smaller for higher $p$ and larger for lower $p$.

\section{Scalar Wave Emission}\label{sec:Emission}

A very interesting question is what happens during mergers. The LIGO/Virgo collaboration has measured multiple black hole mergers, broadly  compatible with the predictions of general relativity \cite{LIGOScientific:2016aoc,LIGOScientific:2020tif}. One might wonder if this puts a further bound on these scalar models. For a canonical scalar the answer is no. Black holes neither source or are affected by scalars. One can picture this as follows: as matter flows towards a black hole horizon it becomes ultra-relativistic, but scalars do not couple to ultra-relativistic matter which have vanishing trace of energy momentum tensor $T$. Furthermore, black holes do not source scalars as this would violate the black hole no-hair theorem \cite{Bekenstein:1995un,Saa:1996aw}; the black hole should not ``remember" that it was formed from material that was coupled to a scalar as opposed to some other material that is not (if we allowed for non-universal couplings). A no-hair theorem is less clear in the presence of a non-canonical scalar with unusual kinetic terms. However, there is some work supporting this in the literature \cite{Hui:2012qt,Graham:2014mda}; this deserves further consideration.

On the other hand, merging neutron stars will source and respond to $\phi$, perhaps until the final stage of black hole formation. During the in-fall stage, the scalar radiation can be appreciable. Here we would like to compute its amplitude and compare to that of general relativity.

\subsection{Effective Metric and Perturbative Scheme}

To make progress, we shall closely follow the very helpful work in Ref.~\cite{Dar:2018dra} (other useful work includes Ref.~\cite{Chu:2012kz}).
In that work, the focus was on Galileon models. But we shall adapt much of their method to our kinetic term. 

We wish to study radiation from a binary system. Since our theory is highly nonlinear, we cannot solve it exactly analytically. So to proceed, we need some perturbative scheme to operate in. The idea is to take the background solution $\phi_0$ about which to expand to be given by the solution from a point source. For a coupling function $f(\phi)=(1-\beta\phi)^2$, the scalar's source (right hand side of Eq.~(\ref{phifull})) becomes
\beq
J_0=\beta\,M_\phi\,\delta^3({\bf x})
\eeq
with $M_\phi=\int d^3x\,(\rho-3\mathcal{P})$. To leading order, we can take a flat spacetime background $g_{\mu\nu}=\eta_{\mu\nu}$ and so the corresponding equation for the background $\phi_0=\phi_0(r)$ is as given before in Eq.~(\ref{gradphisoln}), whose asymptotic solutions are given in Eq.~(\ref{gradphilimits}). Let us suppose we have this function $\phi_0(r)$.

Now in order to analyze the binary, we wish to perturb around this as
\beq
\phi({\bf x},t)=\phi_0(r)+\pert({\bf x},t)
\eeq
and we wish to find $\pert({\bf x},t)$ to leading order. For an orbiting binary system, $\pert$ will oscillate in time and carry scalar radiation out of the system.  To proceed, it is convenient to return to the action, expand to quadratic order in $\pert$, and dispense with the linear term which must vanish by the Euler-Lagrange equations. Ignoring standard gravity, this gives an action for $\pert$ of the form
\beq
S_2=\int d^4x\left[{1\over 2}G^{\mu\nu}_0\partial_\mu\pert\partial_\nu\pert+\pert\,\Delta J\right]
\eeq
where we have an effective metric, evaluated on the background $\phi_0(r)$ solution from the monopole. In spherical co-ordinates, its value is found to be
\bea
&&G_0^{tt}= K_X(X_0)\\
&&G_0^{rr}= -(K_X(X_0)+2X_0K_{XX}(X_0))\\
&&G_0^{\theta\theta}=-{1\over r^2}K_X(X_0)\\
&&G_0^{\varphi\varphi}=-{1\over r^2\sin^2\theta}K_X(X_0)
\eea
with $X_0=-{1\over2}(\phi_0')^2$. 

For a binary system of point masses $M_1$ and $M_2$, the perturbed source $\Delta J$ is
\beq
\Delta J = \beta\left[M_1\,\delta^3({\bf x}-{\bf x}_1(t))+M_2\,\delta^3({\bf x}-{\bf x}_2(t))-M_\phi\,\delta^3({\bf x})\right]
\eeq
with $M_1+M_2=M_\phi$. Note that the integral of $\Delta J$ over space vanishes, so the perturbation $\pert$ is not sourced by the monopole. The dipole is $\int d^3x\,{\bf x}\,\Delta J=\beta(M_1\,{\bf x}_1(t)+M_2\,{\bf x}_2(t))$, which for non-relativistic sources is the total momentum, and hence it vanishes in the center of mass frame. So to leading order, $\pert$ is sourced by the quadrupole moment, as is familiar in general relativity. 

From varying the above action with $d^4x=dt\,dr\,d\theta\, d\varphi \,r^2\sin\theta$, the equations of motion take the form
\beq
\Box_0\pert=\Delta J
\eeq
where we have introduced a modified box operator on the background defined by
\beq
\Box_0\equiv K_X(X_0){\partial^2\over\partial t^2}-{1\over r^2}{\partial\over\partial r}\left((K_X(X_0)+2X_0K_{XX}(X_0))\,r^2\,{\partial\over\partial r}\right)-{K_X(X_0)\over r^2}\nabla^2_{ang}
\eeq
where $\nabla_{ang}^2$ is the 2-dimensional angular Laplacian on the unit 2-sphere. 

The relevant solution is given by the retarded Green's function
\beq
\pert({\bf x},t) = \int d^4y\,G_R({\bf x},{\bf y},t-t_y)\,\Delta J({\bf y},t_y)
\eeq
where the retarded Green's function is for the modified box operator
\beq
\Box_0 G_R=\delta^4(x-y)
\eeq

In order to solve the above type of wave equation, it is useful to obtain the mode functions $\phi_{lm\omega}$ which obey the source free wave equation $\Box_0\phi_{lm\omega}=0$. Using spherical symmetry, these can be decomposed as
\beq
\phi_{lm\omega}=f_{l\omega}(r)Y_{lm}(\theta,\varphi)e^{-i\omega t}
\eeq
Upon substitution, we see that $f_{l\omega}(r)$ must obey the ordinary differential equation
\beq
\left(-\omega^2+{l(l+1)\over r^2}\right)f_{l\omega}-{1\over r^2 K_X(X_0))}{d\over d r}\left((K_X(X_0)+2X_0K_{XX}(X_0))\,r^2\,{df_{l\omega}\over d r}\right)=0
\eeq

From these mode functions, one can construct the retarded Green's function as
\beq
G_R(x,y)=\Theta(t-t_y) W(x,y)
\eeq
where the Wightman function is
\beq
W(x,y)=\sum_{lm}i\int_0^\infty d\omega \,\phi_{lm\omega}({\bf x},t)\,\phi^*_{lm\omega}({\bf y},t_y)
\eeq
In order to obey $\Box_0 G_R(x,y)=\delta^4(x-y)$, we need to take a time derivative at $t=t_y$ and demand the Wightman function give the spatial delta-function as
\beq
2K_X(X_0)\partial_t W(x,y)\Big{|}_{t=t_y}=\delta^3({\bf x}-{\bf y})
\label{Wightmanderiv}\eeq

For a binary pair of equal masses $M_1=M_2=M_\phi/2$ with orbital radius $R$ and orbital frequency $\Omega$, one can show (see Ref.~(\!\cite{Dar:2018dra})) that the corresponding power output in the quadrupole is
\beq
P_\phi={15\over 8}\beta^2\,\Omega\, M^2f_{22}(R)^2 
\label{PowerScalar}\eeq
where $f_{22}$ means the radial mode function evaluated at $l=2$ and $\omega =2\Omega$. The $l=2$ reflects the quadrupole, and the $\omega =2\Omega$ reflects that after half a period the binary system returns to itself from the point of view of gravity. 

\subsection{Scalar in Linear Regime}

To set the stage, let us begin with the linear regime in which $K(X_0)=X_0$. Then we obtain the standard wave operator
\beq
\Box_0 =\Box = {\partial^2\over\partial t^2}-{1\over r^2}{\partial\over\partial r}\left(r^2\,{\partial\over\partial r}\right)-{1\over r^2}\nabla^2_{ang}
\eeq
and so the corresponding radial mode functions $f_{l\omega}$ must satisfy
\beq
\left(-\omega^2+{l(l+1)\over r^2}\right)f_{l\omega}-{1\over r}{df_{l\omega}\over dr}-
{d^2 f_{l\omega}\over dr^2}=0
\label{RadialModeFunctionLinear}
\eeq
This is a standard form of the Bessel differential equation. The relevant solutions that are well behaved at the origin are the Bessel functions of the first kind
\beq
f_{l\omega}(r)={\norm\over\sqrt{r}}\,J_{l+1/2}(\omega\,r)
\eeq
where $\norm$ is a normalization factor. Note that this says that at large radii $\omega\,r\gg 1$, the (real part of) the mode functions scale with position and time as
\beq
\mathcal{R}[\phi_{lm\omega}]\propto {\cos(\omega\, r-\omega\,t+\gamma)\over r},\,\,\,\,\,\,\,(\omega\,r\gg 1)
\eeq
(where $\gamma$ is a phase). This is the standard asymptotic behavior of radiation propagating at speed $c=1$. 

Inserting this into the condition Eq.~(\ref{Wightmanderiv}), we have
\beq
{2\norm^2\over r}\sum_{lm}\int_0^\infty d\omega\,\omega\, Y_{lm}(\theta,\varphi)Y^*_{lm}(\theta_y,\varphi_y)J_{l+1/2}(\omega\,r)J_{l+1/2}(\omega\,r_y)=\delta^3({\bf x}-{\bf y})
\eeq
Integrating over $\omega$ gives a delta function in radius, then summing over $lm$ gives a delta function in angle. So this matches with a normalization factor of $\norm=1/\sqrt{2}$. 

We insert this into Eq.~(\ref{PowerScalar}) which means evaluating $f_{22}$ at $r=R$. For low velocities, we are in the small $\Omega\,r$ regime. In this regime, we readily expand to leading order, and find
\beq
P_\phi = {4\over15\pi}{\beta^2 M_\phi^2\,\Omega\over R}\left(\Omega\,R\right)^5
\label{PowerCan}\eeq
This is a factor of $\beta^2\mpl^2/3$ (times $M_\phi^2/M^2$, which is usually $\approx1$) smaller than the spin 2 result 
\beq
P_{GR}= {12\over15\pi}{M^2\,\Omega\over \mpl^2\,R}\left(\Omega\,R\right)^5
\label{GRpower}\eeq
 Since we must already assume $\beta\leq 2.5\times 10^{-3}/\mpl$ to pass light bending solar system tests, then a canonical scalar has power that is $P_\phi\leq2\times 10^{-6}\,P_{GR}$, which is very small. However, we now wish to analyze the nonlinear theory to see if this can be significantly enhanced.

\subsection{Scalar in Nonlinear Regime}

Let us now go to the opposite limit in which we are deeply in the nonlinear regime. For the monopole background, $K(X_0)$ is approximated as in Eq.~(\ref{Kapprox}) and the gradient of $\phi_0$ is given by the lower expression in Eq.~(\ref{gradphilimits}). Hence we have
\beq
K_X(X_0)=\left(r\over r_*\right)^q,\,\,\,\,\,\,X_0\,K_{XX}(X_0)=(p-1)\left(r\over r_*\right)^q
\label{KXexp}\eeq
This gives the following wave operator
\beq
\Box_0\equiv \left(r\over r_*\right)^q{\partial^2\over\partial t^2}-(2p-1){1\over r^2}{\partial\over\partial r}\left(\left(r\over r_*\right)^q r^2\,{\partial\over\partial r}\right)-\left(r\over r_*\right)^q{1\over r^2}\,\nabla^2_{ang}
\eeq
So now the radial mode functions satisfy the following update from Eq.~(\ref{RadialModeFunctionLinear}) to
\beq
\left(-\omega^2+{l(l+1)\over r^2}\right)f_{l\omega}-{1\over r}{df_{l\omega}\over dr}-
(2p-1){d^2 f_{l\omega}\over dr^2}=0
\label{RadialModeFunctionNonLinear}
\eeq
We note that the first two terms here are standard. However, the final term now picks up a prefactor of $2p-1$. For $p<1/2$ this term changes sign, which is a reflection of hyperbolicity breakdown, as we already mentioned in Section \ref{HypSuperConstraints}. 

Again the relevant solutions that are well behaved at the origin are Bessel functions of the first kind. These are now
\beq
f_{l\omega}(r) = {\norm\over\sqrt{r^{1+q}}}\,J_\nu(k\,r)
\eeq
where the order of the Bessel function $\nu$ and its wavenumber $k$ are given by
\bea
&&\nu = {1\over2}\sqrt{1+(2+q)(2l+2l^2+q)} \\
&&k={\omega\over\sqrt{2p-1}}\label{wavenum}
\eea
Let us again comment on the large radii behavior $k\,r\gg 1$. The (real part of) the mode functions now scale with position and time as
\beq
\mathcal{R}[\phi_{lm\omega}]\propto{\cos(k\,r-\omega\,t+\gamma)\over r^{1+q/2}},\,\,\,\,\,\,\,(k\,r\gg 1)
\label{ModeFunctionLimit}\eeq

There are two interesting points about this. Firstly, recalling the wavenumber expression Eq.~(\ref{wavenum}), we see that the speed of the radiation is
\beq
c_s=\sqrt{2p-1}
\eeq
for the wave in the nonlinear regime (before relaxing to $c_s=1$ in the linear regime). So if $p>1$ then we have superluminal radiation, for $p=1$ it is luminal, and for $p<1$ it is subluminal. This reinforces the point already made in Section \ref{HypSuperConstraints}. 

Secondly, the scaling of $\propto 1/r^{1+q/2}$ looks unusual at first sight as it is therefore no longer scaling as $\propto 1/r$ which is standard for radiation. However, let us examine the momentum density. The radial momentum density in our theory is given by
\beq
\Pi_r = -K_X(X)\,{\partial\phi\over\partial t}{\partial \phi\over\partial r}
\eeq
Expanding to quadratic order (and ignoring the linear term as it time averages to zero) we have
\beq
\Pi_r. = -(K_X(X_0)+2X_0K_{XX}(X_0)){\partial\pert\over\partial t}{\partial \pert\over\partial r}
\eeq
In the nonlinear regime, the pre-factor here $(K_X(X_0)+2X_0K_{XX}(X_0))$ scales as $\propto r^q$. So if we combine this with the above mode functions (\ref{ModeFunctionLimit}), we see that the overall scaling is the standard $\propto 1/r^2$ scaling
\beq
\Pi_r\propto {\sin^2(k\,r-\omega\,t+\gamma)\over r^2},\,\,\,\,\,\,\,(k\,r\gg 1)
\eeq
Then to obtain the power, we can use energy-momentum conservation in this $\phi$ sector to obtain the time-averaged power of
\beq
P_\phi=\int d^3x\,\langle\dot\rho_\phi \rangle = -\int d^3x\,\langle\nabla\!\cdot\!{\bf \Pi}\rangle=-\int d^2S\,\langle\Pi_r\rangle
\eeq
where we used the divergence theorem to integrate over a sphere. Since $\langle\Pi_r\rangle$ scales as $\propto 1/r^2$, the time-averaged power through any concentric sphere is the same.

\subsection{Nonlinear Power and Comparison to Tensor Modes}

Using the above results, we again insert into the condition Eq.~(\ref{Wightmanderiv}). This is now modified to
\beq
{2\norm^2\over r\,r_*^q}\sum_{lm}\int_0^\infty d\omega\,\omega\, Y_{lm}(\theta,\varphi)Y^*_{lm}(\theta_y,\varphi_y)J_{\nu}(\omega\,r)J_{\nu}(\omega\,r_y)=\delta^3({\bf x}-{\bf y})
\eeq
Integrating and summing, we see that this matches for a normalization factor of
\beq
\norm={r_*^{q/2}\over\sqrt{2}}
\eeq
To obtain the power, we again insert these results into Eq.~(\ref{PowerScalar}) and focus on low velocities. To leading order, we now find
\beq
P_\phi = {15\over16\,\Gamma[1+\sqrt{25+14q+q^2}]}\,\beta^2 M_\phi^2\left(\Omega\over R\right)\left(r_*\over R\right)^{\!q}\left(\Omega\,R\over c_s\right)^{\!\sqrt{25+14q+q^2}}
\label{PowerNonCan}\eeq

It is simple to check that for $p=1$ ($q=0$) this recovers the canonical result in Eq.~(\ref{PowerCan}). For $1/2<p<1$ ($q>0$), we obtain an interesting new relationship compared to the spin 2 result Eq.~(\ref{GRpower}). We see that on the one hand, we have an {\em enhancement} due to the $(r_*/R)^q$ factor. This in fact is the familiar enhancement that we have from the ratio of forces; see Eq.~(\ref{ForceRatio}). On the other hand, we also have a {\em suppression} by the final factor of $(\Omega\,R)^{\sqrt{25+14q+q^2}-5}$ (there is also a correction from $c_s$ and the Gamma function, but this is typically only an $\mathcal{O}(1)$ correction). The orbital velocity is $v_{\mbox{\tiny{orb}}}=\Omega\,R$, so we can summarize this all as
\beq
{P_\phi\over P_{GR}}\sim {F_\phi\over F_{GR}}\left(v_{\mbox{\tiny{orb}}}\right)^{\sqrt{25+14q+q^2}-5}
\eeq
Hence the ratio of scalar power to the spin 2 power is even further suppressed than the ratio of forces whenever $1/2<p<1$ ($q>0$), which is our regime of interest. Since we expect the ratio of forces to never be much larger than 1, i.e., $F_\phi/F_{GR}\lesssim1$, as this could otherwise potentially disrupt the neutron stars, we obtain an overall suppression by a power of velocity. This implies that the scalar power is negligible in systems well before merging, as the motion is non-relativistic. 
Note that the above result assumes all relevant scales are in the nonlinear regime. Since the wavelengths of radiation can readily be so large as to enter the linear regime, this would not be directly applicable. 

However, as mergers take place, objects and their wavelengths can all enter the nonlinear regime, and the above result can be applicable. Furthermore, orbital velocities increase and reach an appreciable fraction of the speed of light, so this additional velocity suppression is essentially removed. 
So the final stages of mergers may have significant energy output corrections into the scalar. The scalar radiation itself would be very difficult to directly detect as $\beta$ is small. But nevertheless the binary system is losing energy at a rate that is different from general relativity. 

We should also consider the corrections to the tensor power $P_t$ emitted. We previously computed this earlier for the special case of $p=5/6$ with the correction given in Eq.~(\ref{PGRcorrection}). For a general power $p$, we can readily do the case for a circular orbit. We find the result
\beq
P_t=P_{GR}\left(1+6\,\beta^2\mpl^2\left(r_*\over 2R\right)^{\!q}\right)
\label{PowerTensorGen}\eeq
This agrees with Eq.~(\ref{PGRcorrection}) with $\epsilon=0$ and $a=2R$, but generalizes this to any power $q=4(1-p)/(2p-1)$. Generally, we have the scaling here $\delta P_t/P_{GR}\sim F_\phi/F_{GR}$, without any velocity suppression.

Altogether, both the scalar wave emission and the correction to the tensor wave emission can affect the observed signal at LIGO/Virgo or other GW interferometers. A precise comparison to LIGO/Virgo data from neutron star mergers \cite{LIGOScientific:2017vwq,LIGOScientific:2020aai} is beyond the scope of the current paper. This all deserves further investigation.

\section{Discussion}\label{sec:Discussion}

In this work we have developed a class of models which alters the behavior of gravity at macroscopic scales, and yet obeys all hyperbolicity and subluminality constraints, as well as solar system bounds. The coupling to matter remains at the standard bound of $\beta\le 2.5\times 10^{-3}/\mpl$. We found that the scale of new physics that sets the higher dimension kinetic terms is bounded by $\T_1^{1/4}\gtrsim 125$\,eV from Hulse-Taylor binary pulsar precession and a rough constraint of $\T_2^{1/4}\gtrsim 200$\,eV from stability and structure of neutron stars (and both are scaled down by powers of $(\beta/(2.5\times 10^{-3}/\mpl))$ if $\beta$ is lowered further than the current solar system bound). We also computed corrections to the power emitted. We found that the scalar wave emission is even further suppressed by a power of velocity relative to the tensor wave correction; however, it may be important during a merger.

There are many interesting points to discuss. Firstly, one might wonder the plausibility of having a scalar whose leading coupling to matter is very small $\beta\ll 1/\mpl$. If the field were canonical, then under all circumstances the strength of this new force would be weaker than spin 2 gravity. This is not in contradiction with the original form of the ``weak gravity conjecture" \cite{Arkani-Hamed:2006emk}, in which spin 2 gravity is argued to necessarily be the weakest force, as that refers to a comparison to repulsive spin 1 interactions. Other proposals do include extensions to scalars \cite{Palti:2017elp}, though it is less established. An interesting question is whether the non-canonical kinetic term alters this analysis, since the force can rise and potentially become larger than gravity in sufficiently dense environments. Let us also mention that if the scalar has appreciable couplings to very heavy fields it would more easily obey any such constraints (although by giving up universal coupling, it may compromise the plausibility of the lightness of $\phi$). Perhaps of more concern is that in the densest environments, say near a neutron star surface, the field $\phi$ can be super-Planckian in these models, which appears to be in tension with a ``distance conjecture" \cite{Ooguri:2006in} (albeit one should re-analyze in the presence of non-canonical kinetic terms). One possibility is that a UV completion might require the effective theory to fail here and so the bound from neutron star equilibrium may be discarded. This would still allow for other bounds, such as from precession, which remain in the sub-Planckian field regime. 
These are interesting theoretical questions for future consideration.

In this work we studied the interaction between, and scalar/tensor emission from, neutron stars as the densest known environments, as this new force is most important in such regimes. However, it is interesting to know what happens between black holes. As mentioned earlier, canonical scalars obey a no-hair theorem and therefore do not alter the behavior of black holes. Whether this remains valid from scalars with non-canonical kinetic terms is less clear, although some models are found to do so \cite{Hui:2012qt,Graham:2014mda}.
In particular, the singularity at the center of a black hole would probe well beyond the regime of validity of the effective theory, so one might wonder if the new physics alters the scalar no-hair theorem. An interesting direction would be to numerically study a collapsing system with our new interaction and see if and how the new force vanishes as the black hole horizon forms.

The cutoff of $\T^{1/4}_2\gtrsim 200$\,eV from the neutron star bound or the cutoff of $\T^{1/4}_1\ge125$\,eV from the Hulse-Taylor binary are both very low scales compared to those associated with unification physics. We might first mention that they are still much much higher than the standard cutoff in other models, such as massive gravity, which is $\sim m_g^{2/3}\mpl^{1/3}\sim 10^{-13}$\,eV. Furthermore, the length scale associated with these cutoffs are around $1/\T^{1/4}\lesssim 10^{-9}$\,meters. This is a microscopic scale; many orders of magnitude smaller than the size of a neutron star ($\sim 10$\,km) or the Hulse-Taylor binary ($\sim 10^6$\,km) on which we have made use of the effective field theory. So it is plausible that such an effective theory may remain valid in these regimes. 
On the other hand, we are not aware of an explicit UV completion of these models. Since, as we emphasized, they obey standard hyperbolicity and subluminality bounds then a UV completion is plausible. The boundary case of $p=1/2$ is the DBI model, which has a UV completion in the form of a brane bending mode in higher dimension models. But the relevant cases of $1/2<p<1$ remain unclear. In any case, since the energy scale cutoffs are somewhat low, the quantum induced mass for $\phi$ (see discussion in Section \ref{sec:Models})) is anticipated to be extremely small; so this is at least self consistent. We further note that if a small mass is included that leads to a screening on large scales, one may be able to avoid the precession bound (orange in Fig.~\ref{fig:ConstraintPlot}), while the neutron star structure bound (green in Fig.~\ref{fig:ConstraintPlot}) could remain intact (unless the mass was increased considerably).

Further work would be to perform a comparison to LIGO/Virgo data for the observed neutron star mergers \cite{LIGOScientific:2017vwq,LIGOScientific:2020aai}. This may place a further tighter bound on the theory through constraining scalar wave emission or corrections to tensor wave emission. Here we can make use of our results Eqs.~(\ref{PowerNonCan},\,\ref{PowerTensorGen}) (or for $p=5/6$ the more general result Eq.~(\ref{PGRcorrection})), albeit further corrections may be needed in the final phases. 

Another consideration is possible consequences for cosmology. On very large scales, the scalar force relaxes to the linear regime. In this regime, the scalar forces between massive objects is smaller than the standard gravitational force by $F_\phi/F_{GR}=2\beta^2\mpl^2$, which from solar system bounds is already constrained by $F_\phi/F_{GR}\lesssim 10^{-5}$. However, most cosmological measurements are not currently at the level of 5 significant figures precision. Therefore one might anticipate only weaker constraints on large scales. 
Finally, one can consider other high density environments, such as the very early universe before big bang nucleosynthesis, such as above the QCD scale. Here the known material is ultra-relativistic, so on the one hand, it should not couple appreciably to the scalar, as it is conformally coupled. On the other hand, with our assumed non-standard kinetic interaction, there are competing effects as the densities grow. So a close examination is worth considering.

\section*{Acknowledgments}
We thank Jose Blanco-Pillado, Jaume Garriga, Lam Hui, Leah Jacobs, Mark Trodden, Cumrun Vafa, and Alex Vilenkin for helpful discussions. 
M.~P.~H. is supported in part by National Science Foundation Grant No.~PHY-2013953. 
J.~A.~L is supported by a postdoctoral research fellowship under the FCT, I.P. grant No. CERN/FIS-PAR/0027/2021 and partially by the CFisUC project No. UID/FIS/04564/2020.




\end{document}